\documentclass[12pt]{article}
\usepackage[utf8]{inputenc}  
\usepackage[T1]{fontenc}     
\usepackage{textcomp}       
\usepackage{lmodern}         
\usepackage[margin=1in]{geometry}
\usepackage{graphicx}
\usepackage{amsmath}
\usepackage{longtable}
\usepackage{lineno}
\usepackage{xcolor}
\usepackage[style=apa, backend=biber, maxcitenames=2, uniquename=false, doi=true, isbn=false, url=false]{biblatex}
\usepackage{caption}
\usepackage{authblk}
\usepackage{url}
\usepackage{hyperref}

\newcommand{\tcr}[1]{\textcolor{black}{#1}}

\title{Full Dynamical Model (SOCOL:14C-Ex) of $^{14}$C Atmospheric Production and Transport in Application to Miyake Events}
\author[1]{Kseniia Golubenko}
\author[1,3]{Ilya Usoskin}
\author[2]{Edouard Bard}
\author[1]{Sergey Koldobskiy}
\author[4,5]{Eugene Rozanov}
\affil[1]{Space Physics and Astronomy Research Unit and Sodankyl\"a Geophysical Observatory, University of Oulu, Oulu, 90014, Finland \\ \texttt{kseniia.golubenko@oulu.fi}}
\affil[2]{CEREGE, Aix Marseille University, CNRS, IRD, INRAE,
Coll\`ege de France, Technop\^ole de l'Arbois, Aix-en-Provence, 13545, France}
\affil[3]{Institute for Space-Earth Environmental Research, Nagoya University, Furo-cho, Chikusa-ku, Nagoya 464-8601, Japan} 
\affil[4]{Physikalisch-Meteorologisches Observatorium Davos und World Radiation Center (PMOD/WRC), Davos, 10587, Switzerland}
\affil[5]{Ozone Layer and Upper Atmosphere Research Laboratory, Saint-Petersburg State University, Saint-Petersburg, 19850, Russia}
\date{}

\begin{document}

\maketitle 

\section*{Abstract}
Extreme solar particle events (ESPEs) are caused by rare, enormously strong solar eruptions and can produce globally detectable spikes in tree-ring radiocarbon $^{14}$C, known as Miyake events, which serve as precise chronological tie-points and indicators of extreme solar activity. 
After production, radiocarbon is subjected {to the complex carbon cycle}, including large-scale atmospheric transport, which is crucially important for fast and strong Miyake events with highly inhomogeneous $^{14}$C production.
A new 3D dynamical model, SOCOL:14C-Ex, of the radiocarbon atmospheric production and transport is presented here, which can model fast changes in the $^{14}$C atmospheric concentrations with high temporal and spatial resolution.
Precise response curves of $\Delta^{14}$C to a reference ESPE (100$\times$GLE\#69) were computed for various event dates. 
They can be directly applied to analyse Miyake events under different conditions. 
Seven strong events over the past 14 millennia (AD 993, AD 774, 664 BC, 5260 BC, 5411 BC, 7177 BC, and 12351 BC) were analysed by fitting the reference curves to the available annual $\Delta^{14}$C data, identifying the most probable values and confidence intervals of their parameters -- strength, event's date and background level. 
By applying corrections for the geomagnetic and atmospheric (CO$_2$) factors, the strengths of the corresponding ESPEs were assessed. 
The strongest ESPE is confirmed to be that of 12351 BC, while that of AD 774 remains the strongest event during the Holocene.
To conclude, a new tool, based on the radiocarbon atmospheric transport model SOCOL:14C-Ex, is presented to analyse fast changes in the $^{14}$C production.
%===================================
\section*{Introduction}
\label{S:Intro}

Radiocarbon $^{14}$C is a radioactive isotope produced by cosmic rays in the Earth's atmosphere and used to determine the age of carbon-based samples, which forms a basis for precise archaeology.
Initially, it was assumed that the production rate of radiocarbon is constant in time, and the so-called $^{14}$C age represents the true calendar age.
However, it was soon found that the $^{14}$C production rate and its relative concentration vary over time, as modulated by geomagnetic shielding, solar activity, and climate.
To account for that, a calibration-curve approach has been developed as presented in a community consensus IntCal dataset (\cite{reimer20}).
Since these changes are assumed to be slow, the IntCal curve presents a slightly smoothed $^{14}$C dataset with a 5--10-year time resolution.
However, it may sometimes lead to somewhat ambiguous and not very precise dating.
To account for the variable production of $^{14}$C, a carbon-cycle model needs to be applied to relate the production rate to the measured concentrations.
Radiocarbon production and transport in the atmosphere are typically modelled by applying a box model with fixed geometry, which works well for slow changes (e.g., \cite{oeschger74,Bard1997,Buntgen2018}).

As discovered by \cite{miyake12}, {very rarely, radiocarbon concentration can exhibit} strong, well-identifiable spikes in $\Delta^{14}$C corresponding to a nearly instant production of a large amount of $^{14}$C in the atmosphere. 
Such spikes, commonly called 'Miyake events', offer unique time stamps, making the absolute dating of samples covering the corresponding time periods possible (\cite{Heaton2024}).
Accordingly, it is crucially important to study the Miyake events with the highest possible precision and to understand their nature (\cite{Usoskin2023b}).

Here we present a new approach to model fast changes of radiocarbon production and transport in the atmosphere, by utilising a full and precise 3D plus time modelling of the atmospheric dynamics, by using the chemistry-climate SOCOL:14C-Ex model (\cite{Golubenko2025}). 
We describe the model and its application to the known Miyake events to demonstrate its accuracy.

\section*{Miyake events as manifestations of extreme solar particle events}

Miyake et al. (\citeyear{miyake12}) discovered a sudden strong increase of about {12} \textperthousand\ ({subsequent studies refined this estimate of about 18–20 \textperthousand\ (\citeyear{Usoskin2023b})}), in $\Delta^{14}$C corresponding to the year AD 775 (1175 BP).
{The increase was so large that Miyake et al. considered it unlikely to be of solar origin, and it has been hypothesised that an non-solar astrophysical event, such as a nearby supernova, could be responsible for it.}
It can be noted that only gamma-ray emission from a supernova explosion can produce such a short spike (\cite{pavlov13}), while a cosmic-ray signal would have been diluted over centuries or millennia due to the diffusive cosmic-ray transport in the interstellar medium.
However, it was soon demonstrated (\cite{Usoskin2013}) that the enhanced production of $^{14}$C was likely caused by an enormous burst of solar energetic particles, named ESPE (extreme solar particle event).
As discussed by Pavlov et al. (\citeyear{pavlov13}), while a gamma-ray pulse from a supernova or a galactic gamma-ray burst can potentially produce radiocarbon in the Earth's atmosphere, it cannot create a measurable amount of another cosmogenic isotope $^{10}$Be there. 
Clear detection of a $^{10}$Be spike, dated to AD 775 (1175 BP), in both Greenland and Antarctic ice cores (e.g., \cite{Usoskin2013,Mekhaldi2015,Sukhodolov2017}) excluded gamma-rays as a potential source of the Miyake event.
Other {non-solar} sources, such as a cometary impact on Earth (\cite{Liu14}), have also been excluded by a careful analysis of available data (\cite{usoskin_Icarus_15}).
More such cosmogenic isotope spikes {have been discovered since} (e.g., \cite{Miyake2013,OHare2019,Bard2023}), implying that such events are rare but not exceptional, further invalidating their {non-solar} origins.
They are collectively called the `Miyake events' now. 
Presently, five such events are fully confirmed by multi-proxy data, and four remain as candidates (\cite{Usoskin2023b}).
Thus, the current consensus paradigm is that the Miyake events are caused by ESPEs, either single ones or a short sequence of events (\cite{Cliver2022,Usoskin2023b,Usoskin2023a}).
We note that there is no other known feasible source of Miyake events but ESPEs.

The use of different cosmogenic isotopes with different production energy thresholds, viz. $^{14}$C in tree rings, as well as $^{10}$Be and $^{36}$Cl in polar ice cores, for the analysis of ESPEs, {allows for a parametric reconstruction} of the energy spectra of solar energetic particles (SEPs) responsible for them.
Energy spectra of ESPEs, as reconstructed using such multi-proxy analysis, appear fairly similar to those of the directly measured strong SEP events during the space era, but several orders of magnitude stronger (e.g., \cite{Mekhaldi2015,Koldobskiy2023,OHare2019,Paleari2022}).
The similarity of the spectral shapes confirms the solar origin of the Miyake events.
Because of the soft energy spectra of ESPEs production of $^{14}$C is limited mostly to the polar stratosphere, in contrast to the `normal' production by galactic cosmic rays (\cite{Golubenko2022}).

\subsubsection*{Reference ESPE}
\label{SS:reference}

\begin{figure}[t]
    \centering
    \includegraphics[width=0.7\linewidth]{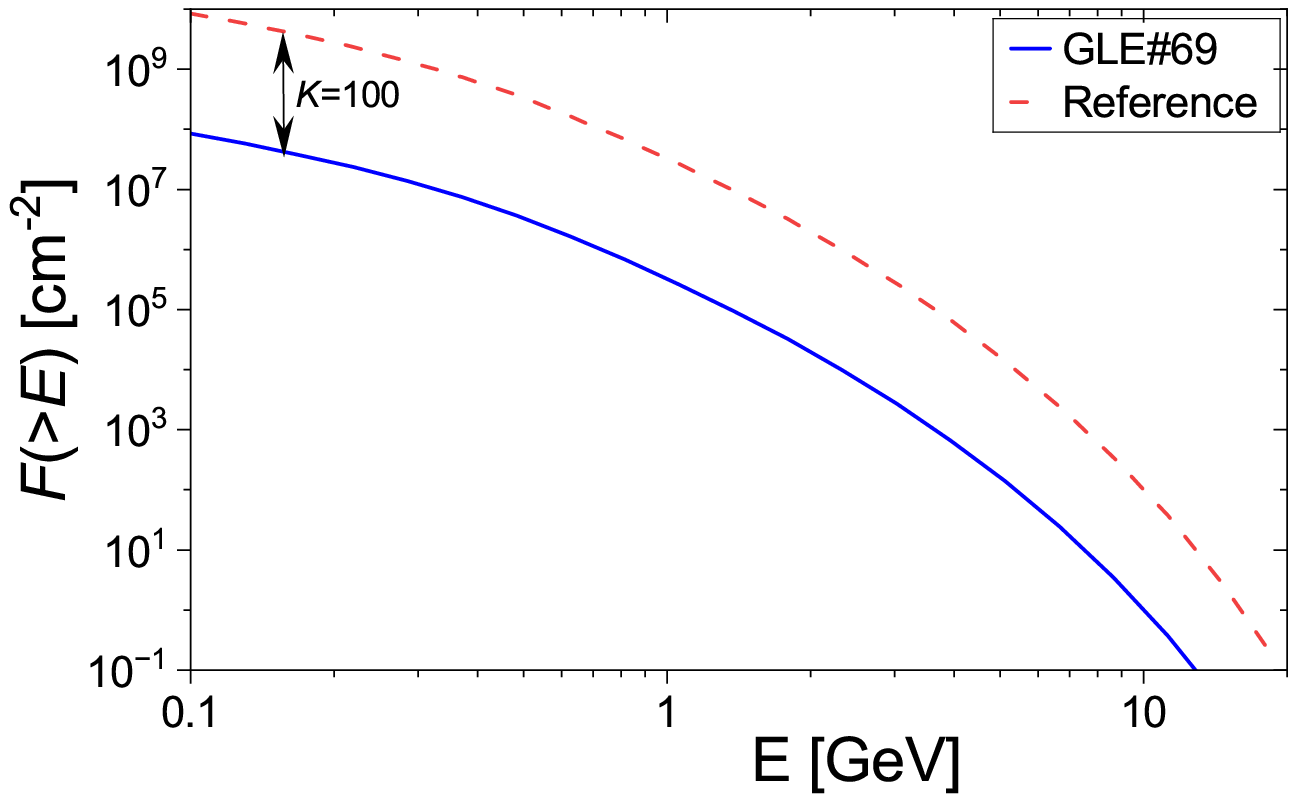}
    \caption{Integral omnidirectional fluence of solar energetic particles, $F(>$$E$) for the GLE~\#69 (20-Jan-2005) as reconstructed from ground-based and space-borne data (blue curve -- \cite{Koldobskiy2021}).
    The red dashed curve is scaled up by a factor $K$=100, representing the reference ESPE spectrum used here.}
    \label{fig:spectrum}
\end{figure}

As a reference ESPE energy spectrum for this work, we used that of a recent hard-spectrum Ground Level Enhancement (GLE) \#69 (20-Jan-2005), which was well measured by a multitude of ground-based and space-based instruments. 
That solar particle event was the second largest ($\approx$4500 \% instant count rate increase at ground-level polar neutron monitors) and one of the best studied events (e.g., \cite{bieber2013}). 
Its energy spectrum was shown to correspond to the ESPE spectral characteristics (\cite{Koldobskiy2023, Mekhaldi2021,Paleari2022}).
The spectral shape was taken following a recent full reconstruction (\cite{Koldobskiy2021}), but the intensity was scaled by a freely adjustable scaling factor $K$, as illustrated in Figure~\ref{fig:spectrum}.
This approach has been validated previously (\cite{Golubenko2025}).
As the reference ESPE, we have considered GLE \#69, scaled up by a factor of $K=100$ (Figure~\ref{fig:spectrum}), which roughly corresponds to the sensitivity of the ESPE detection by cosmogenic isotopes (\cite{Mekhaldi2021,usoskin_1956_20}).
The omnidirectional integral fluence of SEPs with energy above 200 MeV for the reference event is $F_{200} = 2.78\cdot 10^9$ cm$^{-2}$.
We note that the exact spectral shape may only slightly affect the altitude and latitude profile of radiocarbon production, while the total amount of $^{14}$C produced by SEPs is defined by the fluence of SEPs with energy above 200 MeV, $F_{200}$ (\cite{Koldobskiy2022}). 

The modelling of the $\Delta^{14}$C response was performed by the SOCOL:14C-Ex model (see below) for the Holocene conditions with the CO$_2$ concentration set as 287 ppmv, and \textbf {the geomagnetic dipole moment $M=(9.5\pm1.0)\cdot10^{22}$ A m$^2$} as corresponding to the end of the preindustrial epoch (\cite{Panovska2023}).
The total amount of $^{14}$C atoms produced by this reference event in the Earth's atmosphere for this value of $M$ is 3.68$\cdot 10^{26}$, which are produced mostly in the polar stratosphere (\cite{Golubenko2022}).
This would correspond to {an additional} global $^{14}$C production rate of 2.3 atoms cm$^{-2}$ s$^{-1}$, uniformly distributed over one year. 
{This production is superimposed on the normal GCR-related background of about 1.6\,--\,2 atoms cm$^{-2}$ s$^{-1}$ (\cite{Masarik2009,Poluianov2016}), resulting in a total global production rate of about 4 atoms cm$^{-2}$ s$^{-1}$ during the event year, i.e., nearly doubling of the annual $^{14}$C production.}

%\section*{Methods}
\section*{3D dynamical chemistry–climate model SOCOL:14C-Ex}

To model the atmospheric transport of radiocarbon for a very fast and spatially limited $^{14}$C production typical for the Miyake events, we use a full dynamical atmospheric chemistry–climate model SOCOL:14C-Ex, specifically adapted for $^{14}$C tracing (\cite{Uusitalo2024,Golubenko2025}). 
The model couples a general circulation model MA-ECHAM (European Centre/HAMburg climate model), with a detailed atmospheric chemistry module, enabling a fully interactive representation of dynamics and chemistry of the low and middle atmosphere. 
The model's grid is based on horizontal resolution of $2.8^\circ \times 2.8^\circ$ (spectral truncation T42), corresponding to 128 longitude and 64 latitude grid points. 
The vertical domain comprises 39 hybrid $\sigma$–pressure levels, extending from the surface to about 80 km altitude (0.01 hPa), covering the troposphere, stratosphere, and mesosphere.

SOCOL:14C-Ex is a chemistry-climate model of the SOCOL family, built on SOCOLv3 (SOlar Climate Ozone Links, version 3 -- \cite{Stenke2013}) with the AER aerosol microphysics module (\cite{weisenstein1997sulfur,sheng2015sulfur}), the MA–ECHAM5 atmospheric dynamics core (\cite{hommel2011maecham5}), and the MEZON chemistry module (\cite{egorova2003mezon}). 
Aerosol processes and atmospheric chemistry are interactively coupled with circulation, with information exchange between modules every two model hours. 
For $^{14}$C transport, the isotope is treated as a passive gaseous tracer subject to source and sink terms.

The production of $^{14}$C is calculated using the CRAC:14C production model (\cite{Poluianov2016}) for SEPs. 
Radiocarbon produced by galactic cosmic rays (GCRs) is assumed to be a constant background upon which the SEP-related $^{14}$C is added, because the 11-yr solar-cycle related variability in $\Delta^{14}$C is small (about 2 \textperthousand\, -- e.g., \cite{Brehm2021}), and its phase is unknown for ancient periods.

Climatic boundary conditions, such as sea surface temperature, sea-ice cover, and atmospheric CO\textsubscript{2} concentration (\cite{marcott2014carbon}), are chosen to match the exact epoch for the event under consideration (e.g., Late Glacial), and are consistently simulated within the SOCOL framework. 
For the analysed events, full simulations were conducted under realistic atmospheric conditions. 
Each simulation run consisted of six years of spin-up followed by seven years of post-event simulation.

The carbon sink scheme is simplified to include only biospheric uptake and exchange with surface ocean waters, parameterised according to surface albedo, seasonality, and land–ocean distribution (\cite{Golubenko2025}). 
Ice-covered surfaces (albedo $\geq 0.7$) are considered inert. 
Oceanic sinks operate year-round in ice-free regions with an exchange time of 8.29 years (\cite{Guttler2015}). 
Terrestrial sinks operate during the vegetation growth season according to the Leaf Area Index (LAI -- \cite{Verger2023}), defined as: (i) year-round in the tropics, (ii) six months centred on the summer solstice in mid-latitudes, and (iii) absent beyond the polar circles. 

{Return flux from the surface ocean to the atmosphere is neglected during the short time window considered. This modelling approach assumes a steady-state carbon cycle, including the oceanic reservoirs, and is designed to isolate the effect of a hypothesized extra source of very young carbon on the atmospheric $\Delta^{14}$C peak. 
While recent carbon-cycle studies show that the oceanic feedback can influence atmospheric carbon on decadal timescales (e.g., \cite{Miller2025}), transient simulations show that even strong perturbations of ocean circulation translate into relatively modest atmospheric $\Delta^{14}$C changes over the first decades. 
As underlined by \cite{Bard2023}, steady-state estimates based on multi-box models suggest that halving the meridional overturning circulation could increase atmospheric $\Delta^{14}$C by 35–40\textperthousand\ (\cite{Bard1997}), whereas transient simulations over a century yield only about an 8\textperthousand\ increase (\cite{Goslar1995,Hughen1998}). Transients of similar magnitude were obtained with more sophisticated 2D-3D carbon cycle models (\cite{DELAYGUE2003,Singarayer2008}).
Therefore, oceanic feedbacks alone are too slow to explain the magnitude and abruptness of the observed $\Delta^{14}$C peak, and neglecting oceanic return fluxes is justified within the limited temporal scope of the present study. Furthermore, the main problem is that in order to explain even part of the atmospheric $\Delta^{14}$C peak requires an extra source of very young carbon, which is incompatible with oceanic or terrestrial sources (e.g. permafrost).}

The resulting uncertainty (up to 2\textperthousand\, in $\Delta^{14}$C -- see \cite{Golubenko2025}) is taken into account in model-data fitting. 
Long-term decay and deep-ocean overturning of $^{14}$C are not simulated, as their influence on short-term post-event dynamics is negligible.

For robustness, each scenario was simulated as an ensemble of three independent runs with identical boundary conditions. 
The inter-run variability was found to be very small (below 1\% in $^{14}$C concentration, corresponding to $<0.4$\textperthousand\, in $\Delta^{14}$C peak amplitude), and therefore only a single median representative realisation is presented for each event.

In SOCOL:14C-Ex, the simulated output for the near-surface atmosphere is the absolute concentration of $^{14}$C, denoted henceforth as $C$, expressed in $^{14}$C atoms per cubic meter of air (atoms/m$^3$). 
For comparison with observational $\Delta^{14}$C datasets, these values are transformed into per mil (\textperthousand) units (\cite{stuiver77}):
\begin{equation}
\Delta ^{14}{\rm C} = \left({A_{\rm S} \over A_{\rm abs}}-1\right)\cdot 1000,
\end{equation}
where $A_{\rm abs}=226$ Bq/kgC and $A_S$ denotes the specific activity of radiocarbon in near-surface air. 
Assuming the standard air density at sea level, $\rho = 1236$ g m$^{-3}$, this corresponds to $d = 0.123$ grams of carbon per cubic meter of air. 
The number of $^{14}$C atoms per kilogram of carbon is then obtained as $C/d$. 
The computed concentration can be converted to activity using the mean lifetime of $^{14}$C, 
$\tau = 2.62 \cdot 10^{11}$ s as
\begin{equation}
A_{\rm S} = {C\over d\ \tau}\,\, {\rm [Bq/kg C]}.
\label{Eq:A1}
\end{equation}
Finally, the concentration can be directly converted into the corresponding $\Delta^{14}$C value as 
\begin{equation}
\Delta ^{14}{\rm C} = 3.305\cdot 10^{-5}\times {C\over\nu},
\label{Eq:D14}
\end{equation}
where $\nu$ is CO$_2$ concentration (in ppmv) for the period of interest.

\section*{Modelled $\Delta^{14}$C response to an ESPE }

\begin{figure}[t]
    \centering
    \includegraphics[width=0.7\linewidth]{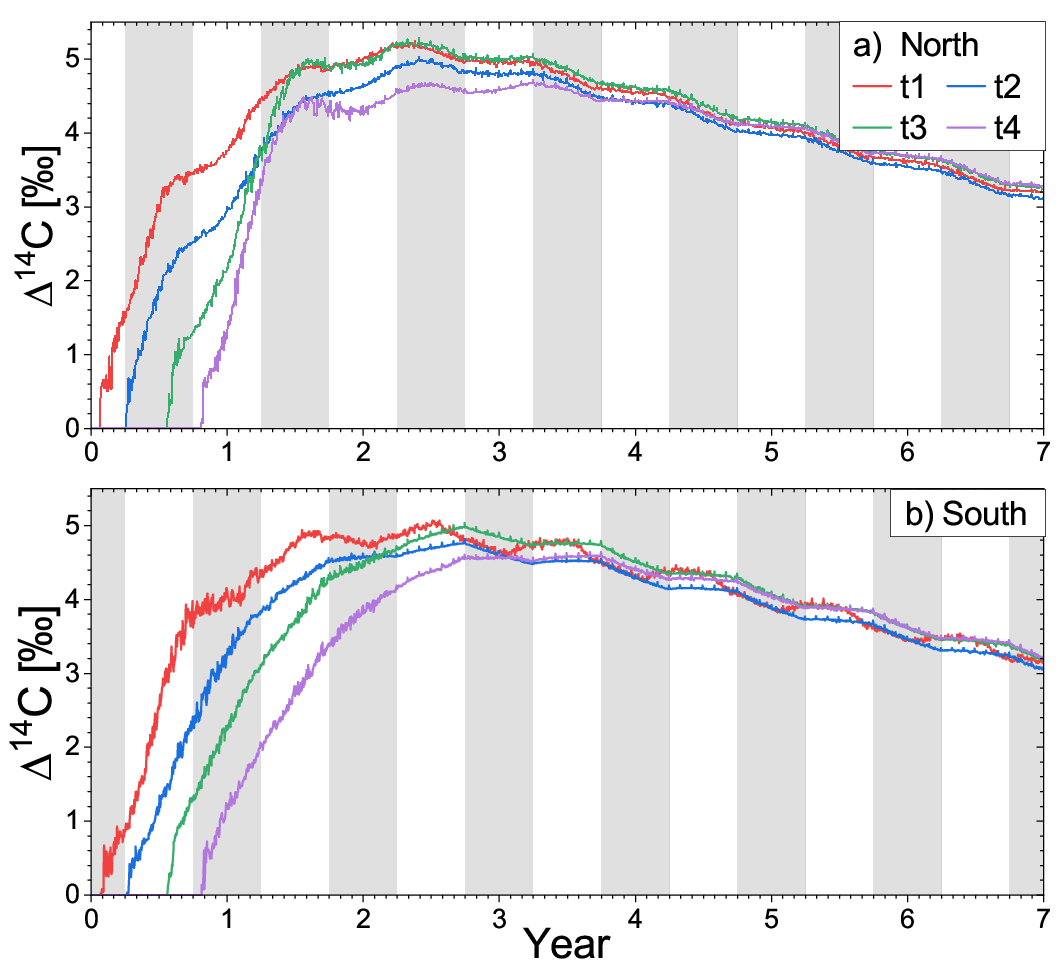}
    \caption{Examples of the time response functions of $\Delta^{14}$C to the reference ESPE (Figure~\ref{fig:spectrum}) for two geographical locations: Central Europe (44.31$^\circ$ N, 5.52$^\circ$ E -- panel a), and Patagonia (41.9$^\circ$ S, 72.67$^\circ$ W -- panel b).
    Different curves correspond to different dates of the ESPE occurrence, as indicated in the legend: 20-Jan, 01-Apr, 20-Jul, and 20-Oct of year zero (1865 in the simulation), denoted as $t1$\,--\,$t4$, respectively.
    Shaded areas approximately indicate the tree growth periods.
    The results are shown with daily resolution, depicting, in particular, meteorological noise on the synoptic scale.}
    \label{fig:daily}
\end{figure}

The modelled daily $\Delta^{14}$C response to the reference ESPE is exemplified in Figure~\ref{fig:daily} for four dates of the event occurrence: 20-Jan, 01-Apr, 20-Jul, and 20-Oct, denoted as $t1$\,--\,$t4$, respectively.
The responses are shown for the end of the preindustrial era (zero time corresponds to 01-Jan-1865, CO$_2$ concentration $\nu=287$ ppmv), for two nearly antipodal locations: Central Europe (44.31$^\circ$ N, 5.52$^\circ$ E), and Patagonia (41.9$^\circ$ S, 72.67$^\circ$ W).
For the computations, the production was modelled as a $\delta$-function (instant production).
In case of prolonged production, {e.g., a series of events separated by days\,--\,months, as produced by a long-living solar active region}, a superposition of response functions can be used.

The modelled $\Delta^{14}$C profiles depict a weak annual cycle caused by the stratosphere-troposphere exchange in the spring season, and the seasonality of the carbon sinks.
The more pronounced annual cycle in the Southern Hemisphere reflects the stronger and more stable Antarctic polar vortex compared to the Northern Hemisphere. 
Such a strong vortex retains the isotopic signal longer at high latitudes and leads to sharper `pulsations' in the exchange with lower latitudes. 
In the Northern Hemisphere, the circulation is more ragged, due to frequent stratospheric sudden warmings (SSWs) and vortex breakdowns, leading to better smoothing of the annual cycle in \(\Delta^{14}\)C. 
It can also be seen that the response is slightly weaker in the Southern Hemisphere, because of the larger sink area in the ocean.

\begin{figure}[t]
    \centering
    \includegraphics[width=\linewidth]{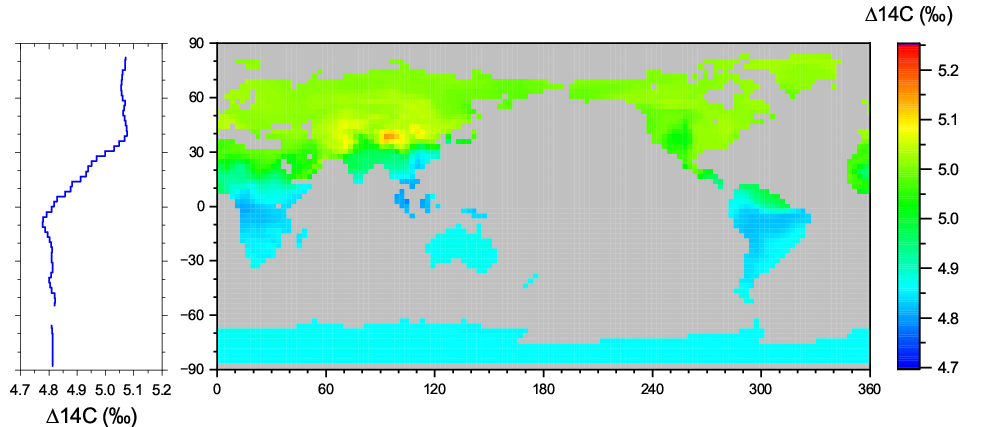}
    \caption{Geographical distribution of modelled near-ground overland $\Delta^{14}$C values caused by the reference ESPE, which took place on 20-Jan-1865.
    The distribution is shown for the day of 20-Jul-1867.
    The LHS panel depicts the latitudinal zonal (over land) mean.
 }
    \label{fig:maps}
\end{figure}
The above curves were shown for the prescribed geographical locations, but the model results are defined for all locations on Earth.
As an example, Figure~\ref{fig:maps} shows a snapshot of the modelled $\Delta^{14}$C geographical distribution over land for the day of 20-Jul-1867, viz. 2.5 years after the reference event occurred on 20-Jan-1865. 
The distribution appears uneven, with several distinct features.
First, there is a significant hemispheric difference with the $\Delta^{14}$C signal being 0.2\,--\,0.3 \textperthousand\, higher in the Northern Hemisphere.
This is partly related to the fact that mid-July is mid-boreal-summer, after the stratosphere-troposphere exchange event, which brings radiocarbon from the stratosphere down to the surface level.
On the contrary, the polar stratospheric radiocarbon is blocked by the polar vortex in the Southern Hemisphere.
This pattern may change during the boreal winter.
Another interesting feature is the tropical dip in $\Delta^{14}$C, which is caused by the ascending flow of the Brewer-Dobson circulation.
The third feature is regional and observed as a slightly enhanced $\Delta^{14}$C signal in the Himalaya and Gobi desert region, caused by the high altitude and the absence of a carbon sink. 
Such features cannot be modelled by a standard box model and require full dynamic modelling.

The modelled $\Delta^{14}$C response curves are provided with daily resolution, but they need to be compared to the annual $\Delta^{14}$C data measured in tree rings.
Accordingly, we have computed annual $\Delta^{14}$C values averaged over the local tree growth period, separately for Northern and Southern Hemispheres as illustrated by shaded stripes in Figure~\ref{fig:daily}.
The averaging was done individually for each location (see \cite{Golubenko2025} for details).
This accounting for the tree growing season is crucially important to define the timing of the production event as discussed later.

\subsubsection*{A set of the response time profiles}

\begin{figure}[t]
    \centering
    \includegraphics[width=\linewidth]{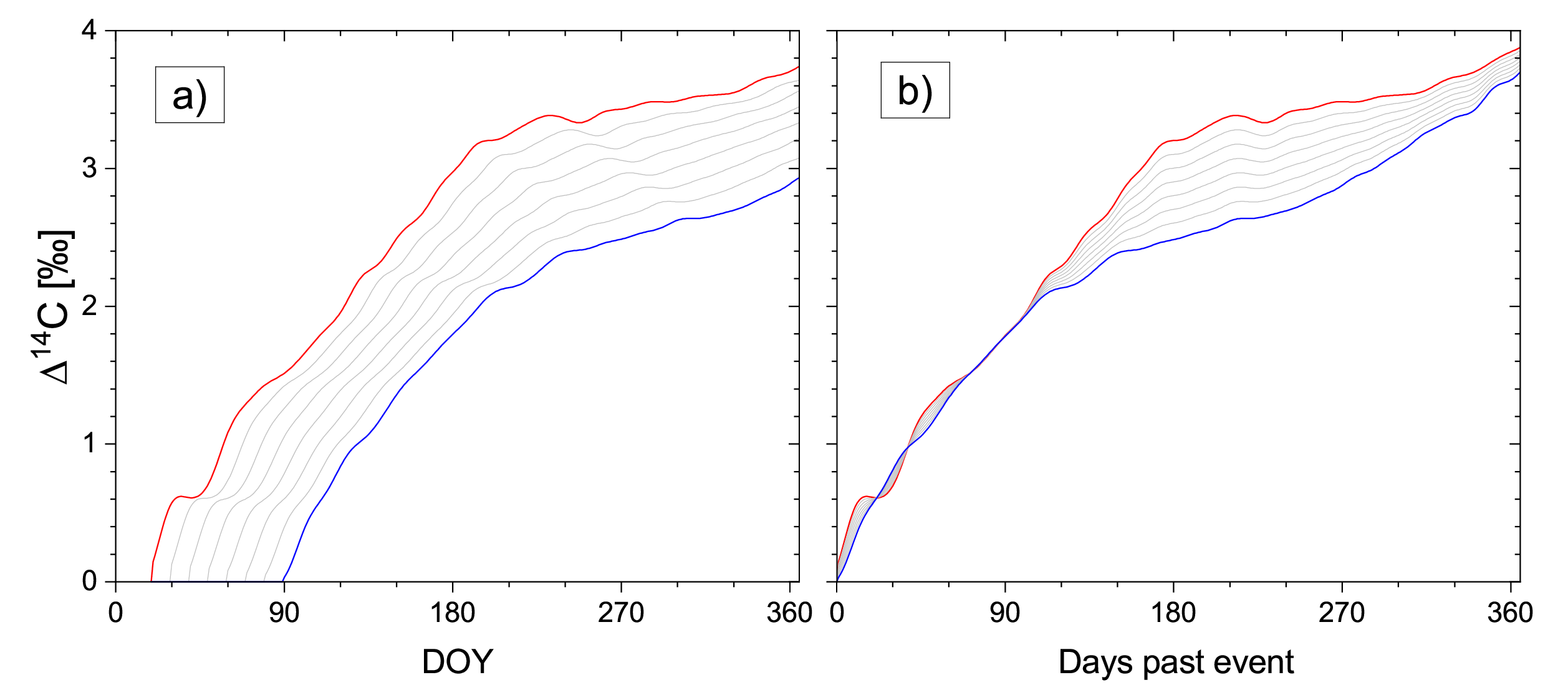}
    \caption{An example of interpolating the response curves for the first year of the ESPE.
    Panel a: full-model calculated responses for 20-Jan (red) and 01-Apr (blue) in the Northern hemisphere -- similar to Figure~\ref{fig:daily}, along with interpolated curves (grey) shown for every tenth day.
    Panel b: Similar to panel a, but all the curves start on the event's date.}
    \label{fig:smooth}
\end{figure}

Only four response curves, corresponding to different seasons, were computed with the full SOCOL model simulation for each hemisphere (Figure~\ref{fig:daily}).
Doing it with a higher cadence would be unfeasible because of the large computational time of $\approx$10 wall-clock days for each curve.
To produce daily response curves, we linearly interpolated between the fully computed curves as illustrated in Figure~\ref{fig:smooth}, assuming that the difference on the short timescale is gradual.
First, the modelled curves were slightly smoothed with a 27-day low-pass filter to remove the meteorological noise -- see red and blue curves in Figure~\ref{fig:smooth}a.
Next, the smoothed modelled curves were joined in time to start on the event day, as shown in Figure~\ref{fig:smooth}b (the event day is day 0).
Then the daily response curves were calculated as a linear interpolation between the nearest fully modelled ones, as shown by grey curves in panel b (every tenth grey curve is shown). 
Finally, the interpolated curves were detouched again to start on the exact event day, as shown by the grey curves in panel a.
These response curves with a daily cadence were further used in the data fitting, as described below.

Interesting in Figure~\ref{fig:smooth}b is that the modelled curves for the 20-Jan and 01-Apr event occurrence dates coincide for the first hundred days and then start diverging.
This can be interpreted so that the first $\Delta^{14}$C response is to radiocarbon produced in the troposphere, with its fast transport for all seasons, on the one hand.
On the other hand, the vertical transport, which brings the stratospheric radiocarbon down, differs between seasons and makes the red curve rise faster due to the fast stratosphere-troposphere exchange in mid-Spring.

\begin{figure}[t]
    \centering
    \includegraphics[width=0.8\linewidth]{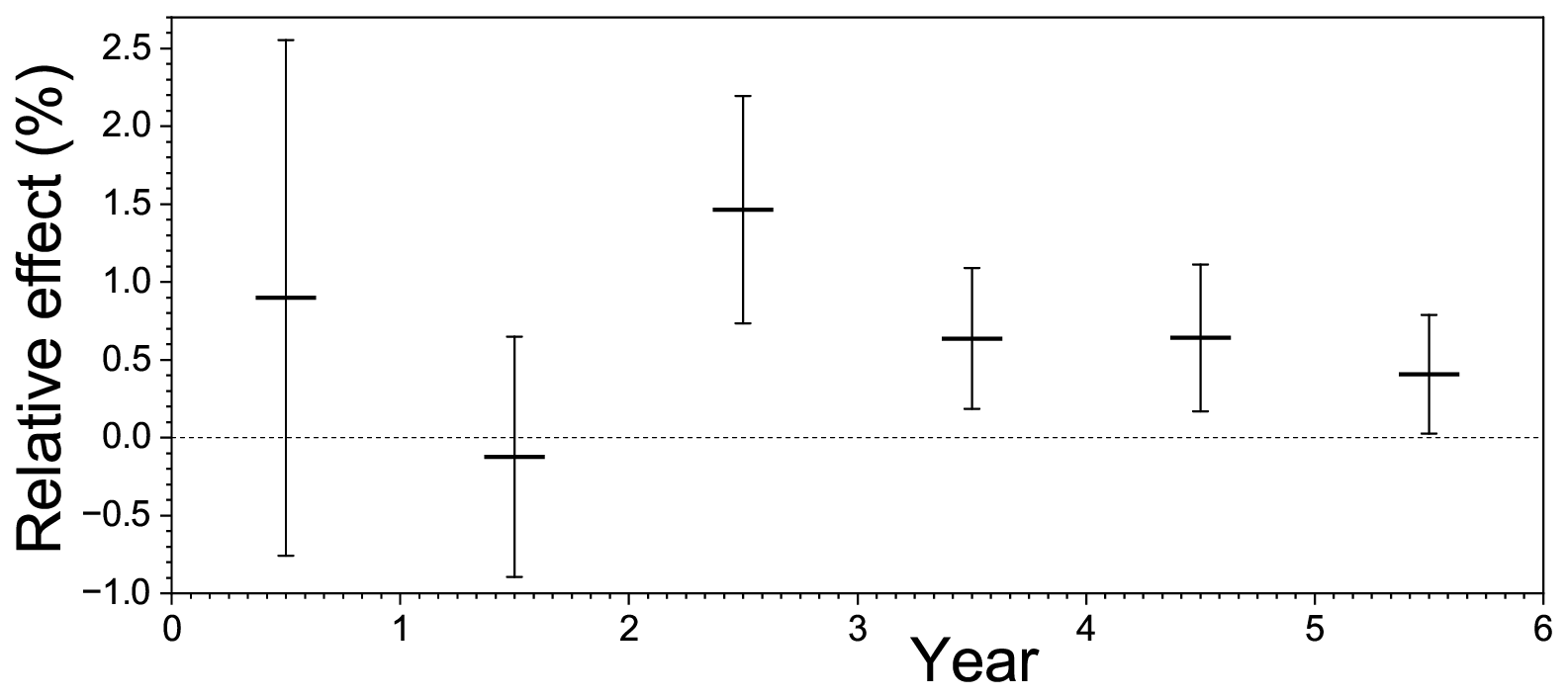}
    \caption{Estimated effect of including a Glacial-type biosphere in the model. The plot shows the percentile difference between the near-ground air $^{14}$C concentrations in Southern Europe, averaged over April--September for years following a reference ESPE occurring at the zero date. The concentrations were computed by the SOCOL:14C-Ex model runs for the Glacial and Holocene vegetation types. The difference is shown as Glacial minus Holocene conditions. The error bars represent the statistical uncertainties between model runs.
    }
    \label{fig:bio}
\end{figure}

The model computations were focused on the Holocene period, but one of the analyzed events, of 12351 BC, took place during the late Glacial period. 
This may affect our analysis in two ways, via the atmospheric transport and the carbon sinks, which could be different during the Glacial.
The former difference was shown to be negligibly small (\cite{Golubenko2025}).
The carbon sink to the ocean is modelled, explicitly considering the ice cover, and thus is adaptive to climate change.
The sink of carbon to the biosphere might be overestimated by the model, since the model considers the Holocene-type vegetation pattern, while vegetation was suppressed during the Glacial (\cite{Prentice1990,Jeltsch2019}).
To evaluate this effect, we performed a modelling of the response curve for $T=20$ DoY, switching off the biospheric sink in the regions identified as extreme deserts (\cite{ADAMS19983}).
The percentile difference between the {reference event's amplitude as reconstructed for} model runs with and without considering the Glacial-type biosphere is shown in Figure~\ref{fig:bio}. 
As seen, the effect is consistent with zero for the first two years, when the produced radiocarbon remains largely in the atmosphere, but {the difference becomes systematically positive yet formally insignificant} from year 3 onward, owing to the reduced biospheric sink.
The effect is negligibly small, about 0.7\% of the production signal, which translates, for the event of 12351 BC, to $\approx$ 0.3\textperthousand.
This exercise can serve as a conservative upper limit for the reduced biosphere's effect.
Accordingly, we neglected this effect in further analysis.

The reference $\Delta^{14}$C response time profiles are presented in Appendix B.

%=================================
\section*{Data analysis}
%-----------

Here, we demonstrate the ability of the new dynamical SOCOL:14C-Ex model to analyse the Miyake event and estimate the parameters of the parent ESPEs, including an assessment of the full uncertainties.

%=======================

\begin{table}[ht]
\centering
\caption{Summary of the analysed Miyake events and ESPEs: year of the ESPE event; atmospheric CO$_2$ concentrations in ppmv (\cite{Indermuehle1999,marcott2014carbon}); VDM $M$ in $10^{22}$~A~m$^{2}$ (\cite{Panovska2023}); $\Delta^{14}$C trend $dC$ in \textperthousand\, year$^{-1}$; scaling of the reference response $A$ (see Table~\ref{tab:ESPE}); background level $C_0$ in \textperthousand \,(Table~\ref{tab:ESPE}); Estimated date of the event $T$ in DoY (Table~\ref{tab:ESPE}); ESPE strength $S$ (Equation~\ref{eq:correction}); and the estimated fluence of SEPs with energy $>$200 MeV $F_{200}$ in $10^9$ cm$^{-2}$.}
\small
\begin{tabular}{c|ccc|ccc|cc}
\hline
{Year} & {CO$_2$} & {$M$} & $dC$ & $A$ & $C_0$ & $T$ & $S$ & $F_{200}$\\
\hline
12351 BC  & 240 & 6.4$\pm$0.8 & -0.03$\pm$0.1 & 5.83$\pm$0.86 & 207$\pm$3.4 & 155$\pm$79  & 3.95$\pm$0.63 & 11$\pm$1.8\\
7177 BC  & 260 & 7.4$\pm$0.6 & 0.04$\pm$0.05 & 3.57$\pm$0.24 & 82$\pm$1.1 & 213$\pm$57 & 2.84$\pm$0.24 & 7.9$\pm$0.7\\
5411 BC  & 265 & 7.3$\pm$0.6 & -0.04$\pm$0.05 & 1.56$\pm$0.25 & 90$\pm$1.0 & 201$\pm$91 & 1.25$\pm$0.21 & 3.5$\pm$0.6\\
5260 BC  & 270 & 7.4$\pm$0.6 & -0.02$\pm$0.1 & 3.73$\pm$0.30 & 94$\pm$1.2 & 182$\pm$54 & 3.10$\pm$0.28 & 8.8$\pm$0.8\\
664 BC   & 275 & 9.1$\pm$0.4 & -0.13$\pm$0.02 & 1.96$\pm$0.19 & 4.3$\pm$0.8 & 21$\pm$60 & 1.83$\pm$0.19 & 5.1$\pm$0.6\\
AD 774  & 285 & 9.3$\pm$0.3 & 0.07$\pm$0.02 & 3.39$\pm$0.19 & -19.7$\pm$0.7 & 133$\pm$40 & 3.35$\pm$0.19 & 9.3$\pm$0.6\\
AD 993  & 285 & 9.0$\pm$0.3 & 0.09$\pm$0.02 & 1.81$\pm$0.13 & -19.9$\pm$0.6 & 54$\pm$50 & 1.56$\pm$0.13 & 4.3$\pm$0.4\\
\hline
\end{tabular}
\label{tab:events}
\end{table}
\subsection*{Miyake events analysed here}

We applied the model and methodology described above to analyse seven Miyake events recorded in radiocarbon for the past 14 millennia, from the Late Pleistocene (12351 BC) to the first millennium AD, as summarised below and in Table~\ref{tab:events}.
We collected available high-resolution data of $\Delta^{14}$C for these events from the literature.
All records are based on annually resolved tree-ring chronologies, some offering earlywood/latewood separation or sub-annual precision.
In such cases, early and late-wood data were subjected to a weighted average to produce annual data.
We did not use the data with a temporal resolution $>$ 1 year.
An inventory of the used datasets and their sources is provided in Appendix A. When dating the event, we use the ESPE year as derived from our analysis, although the onset of the $^{14}$C response can be delayed by about one year (e.g., for AD 774/775). 
The historical date convention (no year zero) is used.

\begin{itemize}
\item {12351 BC (14300 BP) during late glacial period, transition to Holocene}: The highest known $\sim$40 \textperthousand\, rise in \(\Delta^{14}\)C detected in sub-fossil Scots pines from the Italian and French Alps, during deglaciation (\cite{Bard2023}).

\item {7177 BC (9126 BP), Early Holocene, post-Younger Dryas}: One of the strongest radiocarbon anomalies ($\sim$15 \textperthousand) over the Holocene, observed in several high-resolution datasets in the Northern hemisphere (\cite{Brehm2022}).

 \item {5411 BC (7360 BP), Holocene climate optimum}: A strong radiocarbon enhancement ($\sim$ 8 \textperthousand) observed in several high-resolution datasets from the Northern hemisphere (\cite{Miyake2021}).
 
\item {5260 BC (7209 BP), Holocene climate optimum}: One of the strongest Holocene radiocarbon anomalies ($\sim$20 \textperthousand), observed in several high-resolution datasets from the Northern hemisphere (\cite{Brehm2022}).

\item {664 BC (2613 BP), late Holocene}: A strong $\sim$ 10 \textperthousand\, radiocarbon enhancement recorded in several Northern hemisphere trees (\cite{Park2017,Rakowski_Krąpiec_Huels_Pawlyta_Hamann_Wiktorowski_2019,Sakurai2020,Panyushkina2024}).

\item {AD 774 (1176 BP), Early Medieval period}: One of the strongest radiocarbon anomalies ($\sim$18 \textperthousand), observed in several high-resolution datasets in the Northern hemisphere and Southern hemispheres. This is the best-studied ESPE (\cite{miyake12,Usoskin2013,Jull2014,Guttler2015,Park2017,Buntgen2018,Scifo2019,walker2025}).

\item {AD 993 (957 BP), Early Medieval period}: A strong $\sim$10 \textperthousand\, radiocarbon enhancement recorded in several trees in both Northern and Southern hemispheres(\cite{Miyake2013,Buntgen2018}).

\end{itemize}

\subsection*{Data fitting}

First, data from the modelled daily $\Delta^{14}$C curves were averaged over the tree/growth season for each location to represent the annual $\Delta^{14}$C values measured in tree rings as illustrated in Figure~\ref{fig:daily}, similar to \citeauthor{Golubenko2025} (2025).
Thus computed annual modelled $\Delta^{14}$C were compared with the measured data by fitting the annual model curves into the measured data sets and finding the optimal parameters as described below.
The parameters to be determined were: the start date of the event $T$, quantified as the day of year (DoY) of the prescribed year (negative DoY refers to the previous year), leap years were neglected (all years are assumed to have 365 days); the pre-increase level $C_0$ in \textperthousand, which quantifies the background $\Delta^{14}$C level due to GCR; and the amplitude of the event in units of the reference ESPE, $A$.
\begin{equation}
 D^* = A\cdot\mathcal{D}^\dagger(T) + C_0,
    \label{eq:model}
\end{equation}
where $D^*$ denote the measured $\Delta^{14}$C values to be fitted, and $\mathcal{D}^\dagger(T)$ is the modelled response to the reference ESPE, which occurred on day $T$.
The parameters $T$ and $A$ were fitted simultaneously for all datasets for the whole event, while $C_0$ was found individually for each hemisphere.
For each event, we fitted nine annual data points: two for the pre-event years assuming no ESPE signal in the model data, and seven annual points of the event.
If some measured values were missing in a dataset, fewer data points were used in the fit, respectively.

Sometimes Miyake events occur in the background of changing $\Delta^{14}$C level due to the global carbon cycle, as, e.g., for the event of 664 BC (\cite{usoskin_AA_25}).
The trend may distort the fitting and needs to be corrected for. 
We have evaluated the linear trend $dC$ in $\Delta^{14}$C around the events as follows: the measured $\Delta^{14}$C data for 20 years before the event start date and years 20\,--\,40 past the event were fitted by a linear trend considering data uncertainties.
The data were used as: annual for the past three millennia (\cite{Brehm2021,Brehm2025,Fahrni2020}), and 5-year before that (IntCal20 -- \cite{reimer20}). 
The obtained trend values are shown in Table~\ref{tab:events},
As seen, the events of 664 BC, AD 774 and AD 993 depict a weak but significant trend, while the trend is consistent with zero for other events.
For the fitting, the data were detrended using the $dC$ values from Table~\ref{tab:events}.

To assess the robustness of the procedure, the fits were performed by applying two different approaches -- the fitting based on $\chi^2$ statistics, and the MCMC (Markov Chain Monte Carlo) approach, as described below.

%-------------------
\subsubsection*{Statistical $\chi^2$ method}

The statistical approach is based on the $\chi^2$-statistics.
The modelled curves are assumed to be error-free, with the uncertainties being related to the measured data.
As the merit function for the fit, we considered the $\chi^2$ value defined as
\begin{equation}
\chi^2=\sum_i^N\sum_{j=1}^{k_i}{\left({{D^*_{j} - D_{i,j}}\over\sigma_{i,j}}\right)^2},    
\label{eq:chi2}
\end{equation}
where $D_{i,j}$ and $\sigma_{i,j}$ are the values of $\Delta^{14}$C measured in $i$-th dataset for $j$-th year along with their $1\sigma$ error bars, respectively; $D^*_{j}(T)$ are the corresponding modelled $\Delta^{14}$C values for the $j$th year, defined by Equation~\ref{eq:model}, and $k_i$ is the number of meaningful annual points in the $i$-th dataset ($k$=9 if there are no missing data points).
The modelled curves $\mathcal{D}^\dagger$ and the pre-increase level $C_0$ were considered separately for the northern and southern hemispheres, while the values of $A$ and $T$ were the same for each event over the entire globe.

The parameter values were scanned in a reasonable range, and a set of parameter values, minimising the value of $\chi^2$, viz. $\chi^2_{\rm min}$, was set for each studied event. 
The 68\% confidence intervals of the parameters were defined as bounded by the condition of $\chi^2\leq\chi^2_{\rm min}+3.53$ for three independent parameters, viz. $A$, $C_0$ and $T$.
An example is shown in Figure~\ref{fig:993_chi2}, and the best-fit values and the results are summarised in Table~\ref{tab:ESPE}.

%---------------
\subsubsection*{MCMC approach}

\begin{figure}[t]
    \centering
    \includegraphics[width=0.7\linewidth]{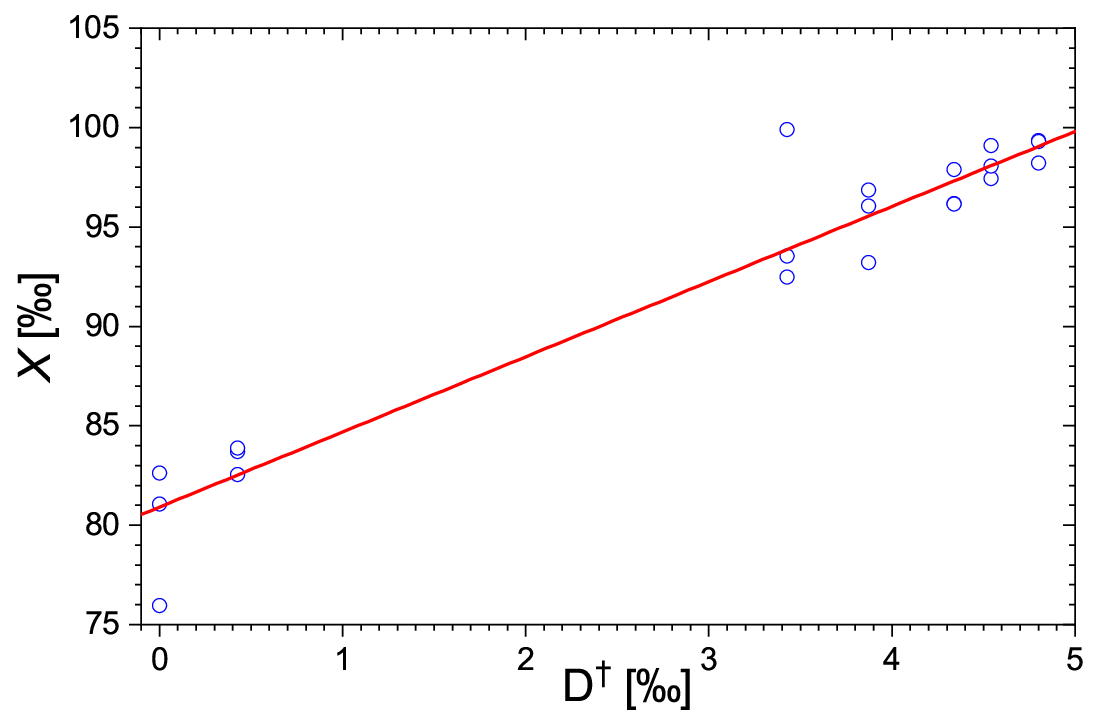}
    \caption{Example of the MCMC determination of the best-fit parameters for the event of 7177 BC (9126 BP).
    Blue dots depict annual values $X$ for one random realisation of three datasets (Y-axis) vs. the reference model curve $\mathcal{D}^\dagger$ for $T$=207 doy.
    The red line is the linear regression $X_{i,j}=3.78\cdot \mathcal{D}^\dagger_j + 80.9$.}
    \label{fig:7176_MCMC}
\end{figure}

The MCMC approach is not based on any specific statistics, but includes all possible uncertainties straightforwardly via randomisation of the datasets.
This was done in the following way.
\begin{enumerate}
\item First, a randomised dataset $X_{i,j}=D_{i,j}+R_{\rm n}\cdot\sigma_{i,j}$ is produced, where $D_{i,j}$ and $\sigma_{i,j}$ are the measured values of $\Delta^{14}$C in the $i$-th dataset for the $j$-th year along with their uncertainties, and $R_{\rm n}$ are normally distributed random numbers with zero mean and unit dispersion.
\item The hypothetical date of the event $T$ is selected that determines the selection of the fitting curves $\mathcal{D}_j^\dagger$, which were considered separately for the two hemispheres.
The randomised dataset $X$ is fit linearly by the fitting curve $D^*(T)$ as illustrated in Figure~\ref{fig:7176_MCMC}.
The best-fit values of $C_0$ and $A$ are found by the least-squares method. 
The RMSE (root mean squared error) value $\epsilon$ was calculated for the fitting.
\item Step 2 is repeated 365 times by scanning, with a daily cadence, over $T$ for one year around the guessed event date. 
The scan can be extended if needed.
The set of $C_0'$, $A'$, and $T'$ providing the absolute minimum of RMSE is fixed.
\item Steps 1\,--\,3 are repeated $N=10000$ times, each time using a new set of $X$ based on a random-number set $R_{n}$.
This yields a distribution of the parameter values $C_0'$, $A'$, and $T'$.
From these distributions, the mean values and the standard deviations can be computed as the estimates for the reconstructed ESPE strength (the value of $A$) and timing $T$, as illustrated in Figure~\ref{fig:MCMC_map}.
\end{enumerate}

\subsubsection*{An example of an analysis for 7177 BC (9126 BP)}

\begin{figure}[t]
    \centering
    \includegraphics[width=0.7\linewidth]{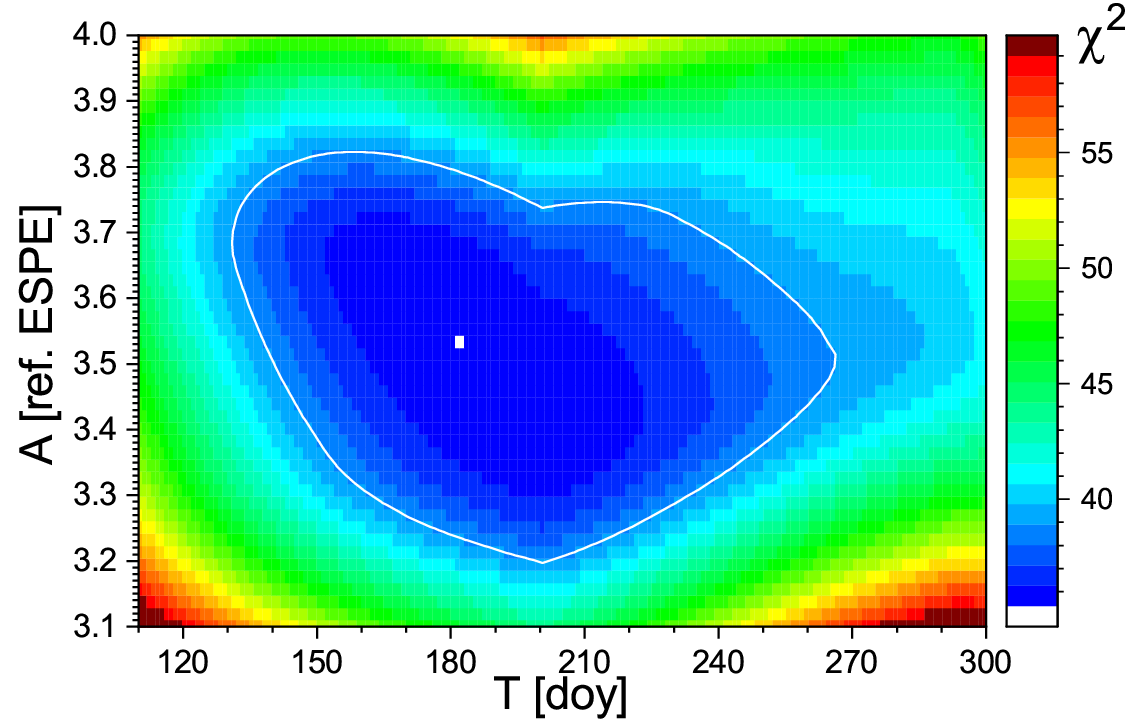}
    \caption{Example of the determination of the best-fit parameters, $T$ (in DoY of 7177 BC (9126 BP)) and scaling factor $A$ for the ESPE of 7177 BC (9126 BP). 
    The colour code represents the value of $\chi^2$ (Equation~\ref{eq:chi2}).
    The best-fit set of parameters ($A$=3.54, $T$=183), corresponding to the minimal value of $\chi^2_{\rm min}$=35.4, is depicted by the white dot, while the white line bounds the 68\% confidence areas ($\chi^2=\chi^2_{\rm min}+3.53$).}
    \label{fig:993_chi2}
\end{figure}

The analysis is illustrated here for the ESPE of 7177 BC (see Table~\ref{tab:ESPE}).
First, we performed the $\chi^2$ analysis as shown in Figure~\ref{fig:993_chi2}, which depicts the dependence of $\chi^2$ on the values of $T$ and $A$ ($C_0$ is fixed at 82 \textperthousand\, for the plot).
The minimum value of $\chi^2_{\rm min}$ is 35.4 (the number of degrees of freedom is 22), corresponding to $A=3.76$ and $T=183$ DoY (white spot in the Figure).
We note that the best fit yields 1.6 $\chi^2$ per degree of freedom, suggesting that the spread of experimental points is broader than defined solely by the formal error bars, or that the used model may be not perfect.
The increment of $\chi^2_{\rm min}$ by 3.53 bounds the 68\% confidence area, as shown by the white line in the Figure.
The best-fit parameters of the ESPE of 7177 BC were found (see Table~\ref{tab:ESPE}) as $A=3.54\pm 0.32$, $T=183\pm 78$ days, and $C_0=82\pm 0.5$ \textperthousand.

\begin{figure}[t]
    \centering
    \includegraphics[width=0.7\linewidth]{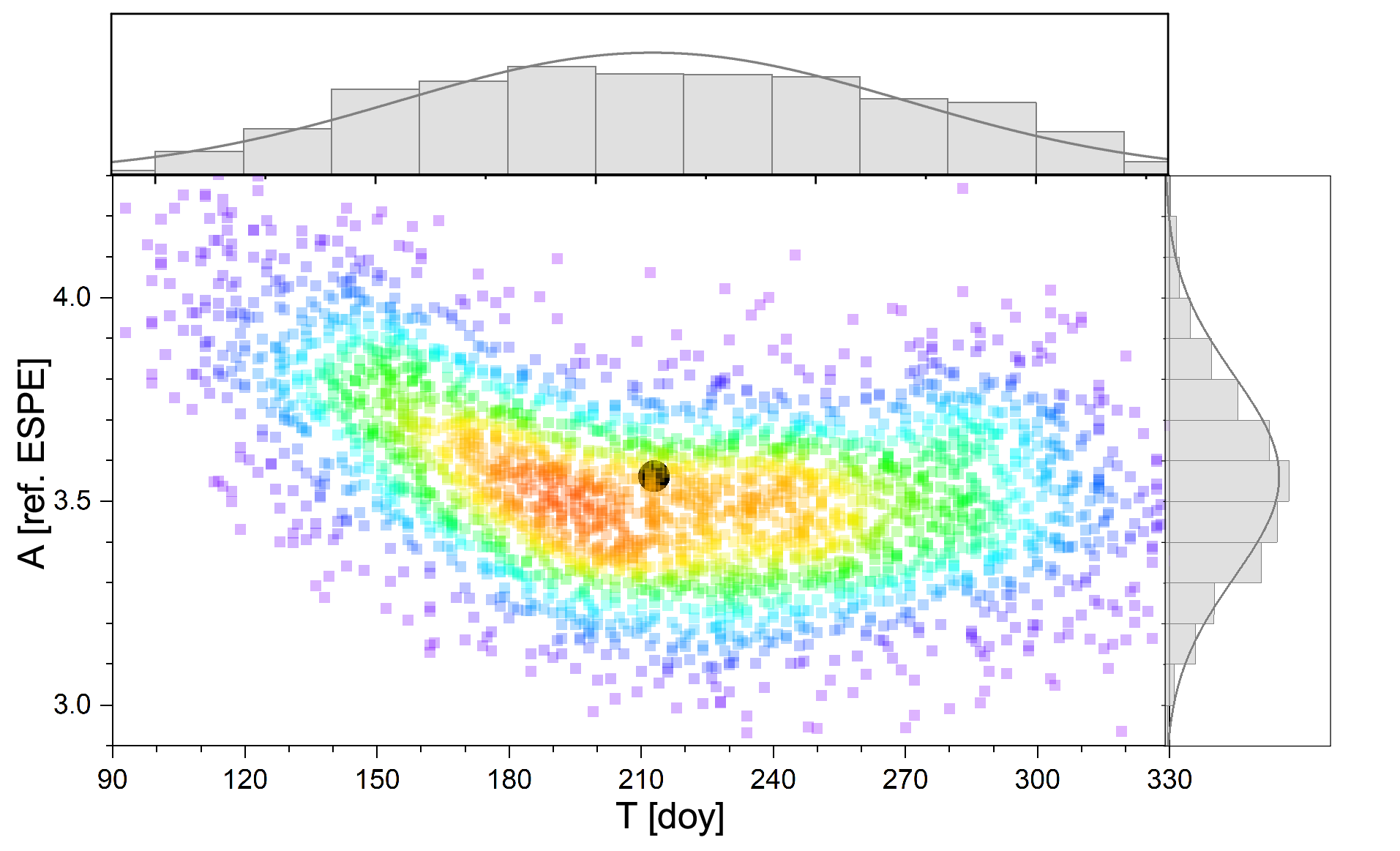}
    \caption{Distribution density of 3000 points in the $A$-vs-$T$ parameter space, for the MCMC analysis of the event of 7177 BC (9126 BP). 
    Each point corresponds to one realisation similar to that shown in Figure~\ref{fig:7176_MCMC}.
    Distributions of the values of $T$ and $A$ are shown on the panels on the top and on the right, respectively.
    The black circle denotes the gravity centre of the distribution ($A$=3.57 and $T$=213).}
    \label{fig:MCMC_map}
\end{figure}

\begin{figure}[t]
    \centering
    \includegraphics[width=0.8\linewidth]{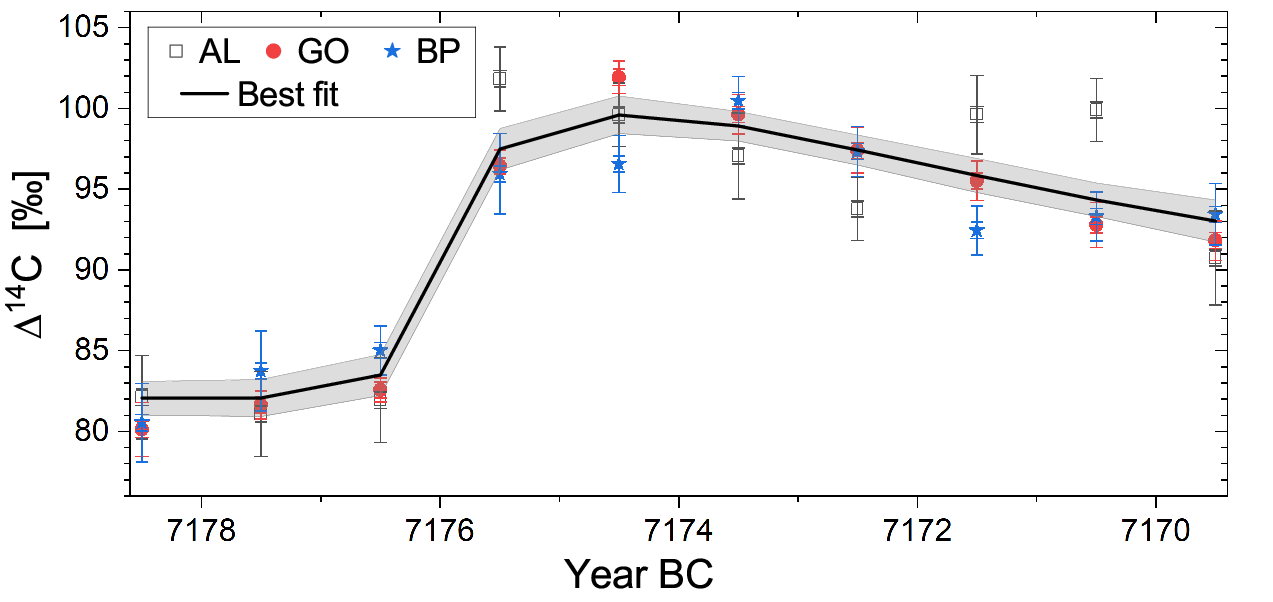}
    \caption{Illustration of fitting of the measured $\Delta^{14}$C datasets (points) for the ESPE of 7177 BC (9126 BP) with the model curves using the MCMC method.
    The datasets are from \cite{Brehm2021}: Alpine Larch (AL), German Oak (GO), and Bristlecone Pine (BP) -- see Table~\ref{tab:events}.
    The best fit with $1\sigma$ model uncertainties (see Table~\ref{tab:ESPE}) is depicted by the solid black line with grey shading.}
    \label{fig:fit}
\end{figure}

An example of the model curve fit by the MCMC method for the same event is shown in Figure~\ref{fig:MCMC_map}.
The central panel depicts the distribution density of the best-fit parameter sets of $A$ and $T$ (values of $C_0$ are not shown).
Distributions of the individual parameter values are shown in the side panels along with the best-fit normal distributions, from which the mean and the standard deviation were obtained for each parameter.
The best-fit parameter set corresponds to the gravity centre of the distribution as denoted by the black dot in the central panel.
The 68\% confidence intervals of the parameter values are defined from the normal distributions (side panels) as $\pm$0.26 and $\pm$57 for $A$ and $T$, respectively, as entered in Table~\ref{tab:ESPE}.
The RMSE value of the fit is $\epsilon\approx$2 \textperthousand , which is comparable to the measurement errors.
The fit is illustrated in Figure~\ref{fig:fit}.

It is important that both $\chi^2$ and MCMC methods, which are totally independent of each other and based on different types of statistics and merit functions, produce very similar results. 
All best-fit parameter values in Table~\ref{tab:ESPE} are close to each other and fully consistent within the 68\% confidence intervals.
For further analysis, we use the results produced by the MCMC method, which is more robust and stable with respect to the measurement error estimates.

The analysis was performed in the same way for all analysed events, with the results summarised in Table~\ref{tab:ESPE}.

\subsection*{Estimate of extreme solar events}

The estimated dates of the events are distributed broadly between winter and late summer for the best-fit values and cover the entire year within the uncertainties.
We have checked with the Kolmogorov-Smirnov test that the distribution of the event dates over the year is consistent with the uniform distribution at the $p-$value $<$0.01.
This suggests that no artificial seasonal bias has been introduced in the analysis.

The strength of the $\Delta^{14}$C responses to the reference event was estimated for the known Miyake events (Table~\ref{tab:ESPE}).
However, the amount of the produced radiocarbon and, thus, the response of $\Delta^{14}$C for the same ESPE may be different, as defined by the conditions during the time of the event: geomagnetic field strength, quantified via the geomagnetic virtual dipole moment (VDM) $M$; the content of CO$_2$ in the atmosphere $\nu$; and potentially the type of climate.
The climate dependence was shown to be negligibly small, within 1\% of the response, i.e. $<$0.4 \textperthousand\, even for the strongest event (\cite{Golubenko2025}).
The other two effects cannot be neglected and must be properly accounted for to relate the strength of the Miyake events in $\Delta^{14}$C to the strength of he corresponding ESPE.

The correction for the CO$_2$ level $\nu$ is a simple linear scaling, as defined by Equation~\ref{Eq:D14}.
The higher the CO$_2$ concentration $\nu$ is, the smaller is the $\Delta^{14}$C response to the same ESPE.
Correction for the geomagnetic field strength is not linear, but also straightforward (see Figure 5 in \cite{Golubenko2025}): the $\Delta^{14}$C response is scaled with the VDM value $M$ as $M^{-0.55}$, viz., the stronger the geomagnetic field is, the smaller is the $\Delta^{14}$C response to the same ESPE.
The adopted values of the VDM and $\nu$ are listed in Table~\ref{tab:events} for the analysed events.
Since the paleomagnetic models are not precise and include uncertainties in the VDM reconstructions, they were included via the MCMC error-propagation approach, into the uncertainties of the final ESPE strength estimates $S$:
\begin{equation}
 S = A\cdot \left({M\over M_0}\right)^{0.55} \cdot {\nu\over\nu_0},
    \label{eq:correction}
\end{equation}
where $M_0=9.5\cdot 10^{22}$ A m$^2$ and $\nu_0=287$ ppmv correspond to the VDM and CO$_2$ concentration for which the reference response was computed.

The reconstructed strengths of ESPEs, $S$, along with their 68\% confidence intervals, are listed in the right-hand-side block of Table~\ref{tab:events}, as well as the corresponding $F_{200}$ fluences of SEPs.
As seen, the correction for geomagnetic and CO$_2$ factors modifies the ranking of the events significantly.
For example, the 12351 BC (14,300 BP) Miyake event was about 72\% higher than that of AD 774, but the corresponding ESPE was only 18\% stronger, in full agreement with the results by \citeauthor{Golubenko2025} (2025).
This is explained by two factors, both enhancing the $\Delta^{14}$C response, viz. weaker geomagnetic field and a smaller CO$_2$ concentration (Table~\ref{tab:events}) -- see \citeauthor{Golubenko2025} (2025) for more details.
While the Miyake event of AD 774 was only the third highest for the Holocene, the corresponding ESPE was found, after the correction, to be the strongest one over the Holocene and the second strongest in the entire record. 
We note that the $F_{200}$ fluence for each ESPE is at least an order of magnitude stronger than the total $F_{200}$ fluence of SEPs, $0.5\cdot 10^9$ cm$^{-2}$, registered over the last three full solar cycles 1984\,--\,2020 (\cite{Raukunen2022}).

\begin{figure}[t]
    \centering
    \includegraphics[width=0.8\linewidth]{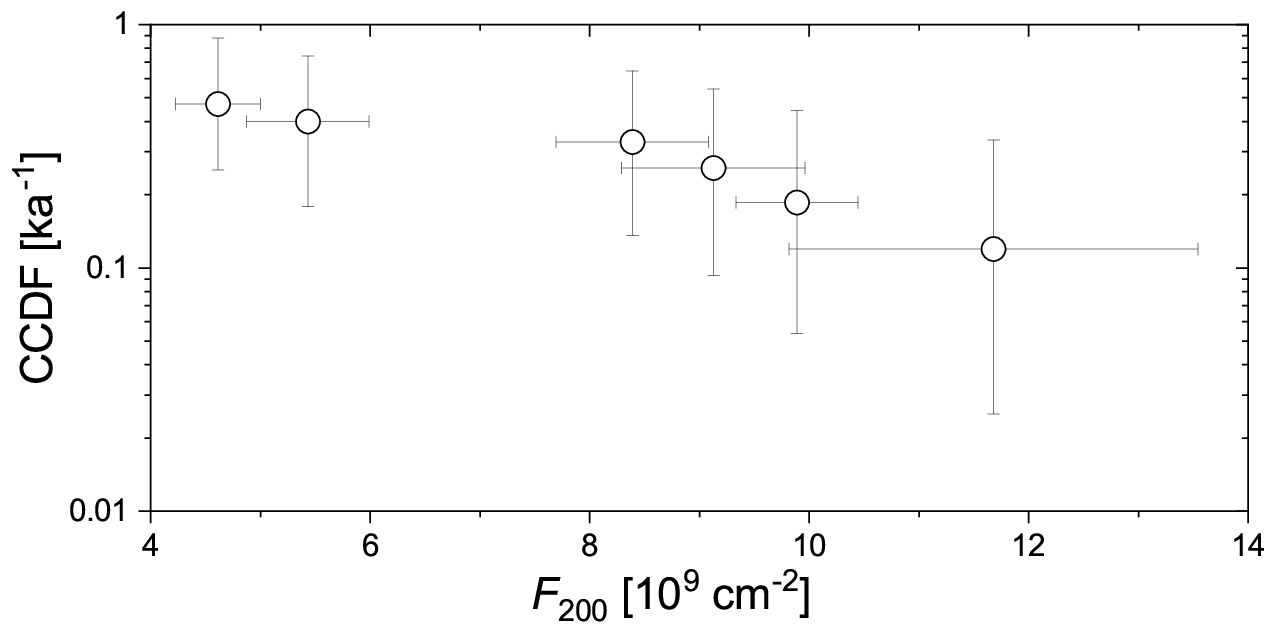}
    \caption{Complementary cumulative distribution function (CCDF) of the occurrence, per millennium, of ESPEs with the $F_{200}$ fluence exceeding a given value, along with the 68\% confidence intervals. 
    The values and error bars of $F_{200}$ are the same as in Table~\ref{tab:events}.
    The CCDF was estimated, along with the confidence intervals, from the Poisson distribution. }
    \label{fig:CCDF}
\end{figure}

Figure~\ref{fig:CCDF} depicts the complementary cumulative distribution function (CCDF) of the occurrence, within a millennium, of an ESPE with the $F_{200}$ fluence of SEP exceeding the given $F_{200}$ value. {The CCDF is computed using confirmed and published events listed in Table~\ref{tab:events}.}
We assumed that all events are independent of each other and thus applied the Poisson statistics to evaluate the CCDF.
The occurrence probability of a clearly detectable ESPE is roughly once in two millennia (68\% confidence interval covers the range from once in three millennia to once per millennium).
The strongest ESPE of 12351 BC (14300 BP) has the occurrence probability of roughly once per ten millennia (ranging from once in three millennia to once in 30 millennia).
In contrast to previous suggestions (e.g., \cite{Cliver2022,Usoskin2023a}), the distribution shows no clear sign of a roll-off at the highest fluences, indicating that the Sun, probably, has not yet reached its limit in producing ESPEs.
However, the present data don't make it possible to make a definite conclusion about that.
\begin{table}[ht!]
    \centering
    \small
    \setlength{\tabcolsep}{4pt}
    \caption{Best-fit parameters, viz. $A$, $T$ [DoY], and $C_0$ [\textperthousand],   along with their 68\% confidence intervals, for the seven ESPEs, analysed here, by two methods -- $\chi^2$ and MCMC.
    Shown are also the values of the merit functions $\chi^2_{\rm min}$ and RMSE $\epsilon$.
    The last column depicts the number of the fitted datapoints $\mathcal{N}$.}
    \begin{tabular}{c|cccc|ccccc}
    \hline
    ESPE & \multicolumn{4}{c|}{$\chi^2$-method} & \multicolumn{5}{c}{MCMC-method}\\
    year & $A$ & $T$ & $C_0$ & $\chi^2_{\rm min}$ & $A$ & $T $ & $C_0$ &$\epsilon$& $\mathcal{N}$\\  
    \hline
    12351 BC & 5.83$\pm$1.13 & 127$\pm$107 & 207$\pm$2 & 13.4 & 5.83$\pm$0.86 & 155$\pm$79 & 207$\pm$3.4 & 3.8 & 14\\
    7177 BC & 3.54$\pm$0.32 & 183$\pm$78 & 82$\pm$0.5 & 35.4 & 3.57$\pm$0.24 & 213$\pm$57 & 82$\pm$1.1 & 2.0 & 22\\
    5411 BC & 1.52$\pm$0.35 & 143$\pm$160 & 90$\pm$0.5 & 16.2 & 1.56$\pm$0.25 & 201$\pm$91 & 90$\pm$1 & 2.1 & 21\\
    5260 BC & 3.74$\pm$0.28 & 144$\pm$51 & 93$\pm$0.5 & 50.7 & 3.73$\pm$0.30 & 182$\pm$54 & 94$\pm$1.2 & 2.4 & 23\\
    664 BC & 2.04$\pm$0.21 & 34$\pm$47 & 4.3$\pm$0.5 & 120 & 2.03$\pm$0.19 & 34$\pm$56 & 4.4$\pm$0.8 & 2.8 & 35\\
    AD 774 &3.3$\pm$0.07 & 133$\pm$15 & -19.5$\pm$0.4 & 571 & 3.32$\pm$0.19 & 131$\pm$38 & -20$\pm$0.8 & 3.5 &308\\
    AD 993 & 1.56$\pm$0.09& 35$\pm$30 & -20$\pm$0.5 & 212 & 1.56$\pm$0.13 & 46$\pm$55 &-20$\pm$0.6&2.8&116\\
    \hline
    \end{tabular}
    \label{tab:ESPE}
\end{table}

\section*{Summary and conclusion}

The main results of this work can be summarised as follows.
\begin{itemize}
\item
A brand-new dynamical 3D model of the atmospheric transport of radiocarbon is presented.
The model SOCOL:14C-Ex belongs to the SOCOL model family of chemistry-climate models and makes it possible, for the first time, to model $^{14}$C concentrations in the atmosphere with high temporal and spatial resolution.
\item
Precise calibration curves of the $\Delta^{14}$C response to a reference extreme solar particle event were computed with a daily resolution for two typical mid-latitude locations for the Northern and Southern hemispheres, as a function of the event date, to account for the geographical and seasonal patterns.
These calibration curves (see Appendix B) can be directly applied to analyses of other Miyake events, as described here.
\item
The SOCOL:14C-Ex model was applied to an analysis of seven strong Miyake events over the past 14 millennia by fitting the calibration reference curves to the available annual $\Delta^{14}$C data.
To secure the robustness of the results, two independent fitting methods were applied, the statistical $\chi^2$ and the MCMC ones, and the results were found fully consistent between them.
The corresponding Miyake events' parameter sets -- scaling $A$, most probable event date $T$, and the pre-event background $C_0$ (Equation~\ref{eq:model}) -- were obtained along with their confidence intervals, as listed in Table~\ref{tab:ESPE}.
\item
The dates of the events were identified, which cover, within uncertainties, all seasons, as expected.
A small insignificant preference for the Spring-Summer season may be related to the higher sensitivity of the Northern Hemisphere trees to such events.
However, the proposed event dates are fully consistent with the uniform distribution of the event throughout seasons.
\item 
By applying corrections for the geomagnetic and atmospheric (CO$_2$) factors, the strength of the ESPEs, responsible for the analysed Miyake events, has been assessed. 
The strongest ESPE is confirmed to be that of 12351 BC, which was 18\% stronger than that of AD 774, which was the strongest event during the Holocene.
\end{itemize}

To conclude, a new tool, based on the radiocarbon atmospheric transport model SOCOL:14C-Ex, is presented to analyse fast changes in the $^{14}$C production when the classical quasi-steady box models may not be sufficiently accurate.

\section*{Acknowledgments}
This work was partly supported by the European Research Council Synergy Grant (project 101166910), Research Council of Finland (projects 330063 and 354280), and by the European Union’s Horizon Europe program under grant agreement 101135044 SPEARHEAD. EB acknowledges support from the MARCARA ANR project. ER acknowledges the support of SPbU (research project ID 124032000025-1). The ISSI Teams 510 (SEESUP, led by F. Miyake and I. Usoskin) and 23-585 (REASSESS, led by A. Mishev) are acknowledged for stimulating discussions.

\section*{Author Contributions}
The idea and concept of this work were discussed and developed by I.U., E.B. and E.R. 
Model development and computations were carried out by K.G., while the fitting and analysis were performed by I.U. and S.K. 
Data was provided by E.B., and S.K. collected the datasets from open sources. All authors contributed to the discussion of the results and their presentation in the manuscript.

%\bibliographystyle{plainnat}
%\bibliography{yourbibfile}

\section*{Appendix A. Inventory of the Radiocarbon datasets}
\label{S:Appendix}

\setcounter{table}{0}
\renewcommand{\thetable}{A\arabic{table}}

The datasets used in this work for the analysis of the Miyake events are listed below. 
\begin{table}
\caption{Metadata of the $\Delta^{14}$C datasets used in this study.
The columns are: the reference to the dataset source; the sample region and name; geographical coordinates and altitude (if available) of the sample location; and the tree \tcr{species}.}
\footnotesize
\begin{tabular}{|l|r|cc|c|}
\hline
Reference & Sample & Coordinates &Altitude, \tcr{m}& Tree \tcr{species}\\
\hline
\multicolumn{5}{|c|}{12351 BC (14300 BP)}\\
\hline
\cite{Bard2023}, Fig.11& Drouz19, FR & 44.\tcr{53}$^\circ$ N, 5.\tcr{87}$^\circ$ E & 860 & Pinus sylvestris \\
% & DR313a, FR & 44.31$^\circ$ N, 5.52$^\circ$ E & 860 m &  \\
% & P305u, IT & 46.15$^\circ$ N, 12.39 E & 1030 m &  \\
\hline
\multicolumn{5}{|c|}{7177 BC (9126 BP)}\\
\hline
\cite{Brehm2022}, Fig,1 & CH & 46.\tcr{57}$^\circ$ N, 8.\tcr{20}$^\circ$ E & -- & Larix decidua \\
 & DE  & 50.05$^\circ$ N, 10.24$^\circ$ E & -- & Quercus robur \\
 & US &37.\tcr{38}$^\circ$ N, 118.\tcr{16}$^\circ$ W& -- & Pinus longaeva \\
\hline
\multicolumn{5}{|c|}{5411 BC (7360 BP)}\\
\hline
\cite{Miyake2021}, \tcr{Table S2}  & US  & 37.3\tcr{8}$^\circ$ N, 118.1\tcr{7}$^\circ$ W & 3094 & Pinus longaeva \\ 
 &  CH & 46.57$^\circ$ N, 8.25$^\circ$ E & 1930 & Larix decidua \\
 & FI$^*$ & 69.26$^\circ$ N, 27.40$^\circ$ E & 199 & Pinus sylvestris \\
\hline
\multicolumn{5}{|c|}{5260 BC (7209 BP)}\\
\hline
\cite{Brehm2022}, Fig.1  & UK & 54.\tcr{47}$^\circ$ N, 6.\tcr{46}$^\circ$ W & -- & Quercus petraea \\
 & US & 37.\tcr{38}$^\circ$ N, 118.\tcr{16}$^\circ$ W & -- & Pinus longaeva  \\
 & CH & 46.\tcr{38}$^\circ$ N, 9.\tcr{87}$^\circ$ E & -- & Larix decidua  \\
 & RU & 67.75$^\circ$ N, 70.06$^\circ$ E & -- & Larix sibirica  \\
 \hline
\multicolumn{5}{|c|}{664 BC (2613 BP)}\\
\hline
\cite{Park2017}, \tcr{SM Table S2} & DE & 49.\tcr{91}$^\circ$ N, 10.\tcr{82}$^\circ$ E & -- & Quercus robur  \\
\cite{Rakowski_Krąpiec_Huels_Pawlyta_Hamann_Wiktorowski_2019}, Tab.1& Grabie, PL & 50.0\tcr{4}$^\circ$ N, 19.\tcr{99}$^\circ$ E & -- & Quercus robur L. \\
\cite{Sakurai2020}, \tcr{SM Table S3} & JP$^\dagger$ & 39.05$^\circ$ N, 140.03$^\circ$ E & -- & Cryptomeria japonica \\
\cite{Panyushkina2024}, \tcr{SM}& Altai, RU & 49.38$^\circ$ N, 89.0$^\circ$ E & 2150 & Larix sibirica \\
 & Yamal, RU & 67.3$^\circ$ N, 70.7$^\circ$ E & 20 & Larix sibirica  \\
 \tcr{\cite{Rakowski2024}, SM Table S1} & \tcr{Grabie, PL$^\dagger$} & \tcr{50.04$^\circ$ N, 19.99$^\circ$ S} & \tcr{--} & \tcr{Quercus robur L.}\\
\hline
\end{tabular}
\\ $^*$ -- Weighted average of several measurements; $\dagger$ -- Early and late wood averaged;
\label{tab:meta}
\end{table}

\begin{table}
\caption*{Table~\ref{tab:meta} continued.}
\footnotesize
\begin{tabular}{|l|r|cc|c|}
\hline
Reference & Sample & Coordinates &Altitude, \tcr{m}& Tree \tcr{species} \\
\hline
\multicolumn{5}{|c|}{AD 774 (1176 BP)}\\
\hline
\cite{miyake12}, \tcr{SM Table S2} & JP$^*$ & 30.\tcr{34}$^\circ$ N, 130.\tcr{52}$^\circ$ E & -- & Cryptomeria japonica \\
\cite{Usoskin2013}, Fig.1 & DE\tcr{$^*$} & 50.0$^\circ$ N, 10.6$^\circ$ E & -- & Quercus robur \\
\cite{Jull2014}, \tcr{SM Table S1} & US & 37.77$^\circ$ N, 118.44$^\circ$ W & \tcr{3539} & Pinus longaeva \\
& \tcr{RU} & \tcr{67.52$^\circ$ N, 70.67$^\circ$ E} & \tcr{350} & \tcr{Larix sibirica} \\
\cite{Guttler2015}, \tcr{Table 3} & NZ & 35.\tcr{92}$^\circ$ S, 173.\tcr{79}$^\circ$ E & -- & Agathis australis \\
\tcr{\cite{Rakowski2015}, Table 1} & \tcr{PL} & \tcr{50.05$^\circ$ N,	20.10$^\circ$ E} & \tcr{--} & \tcr{Quercus robur} \\
\cite{Park2017}, \tcr{SM Table  S1} & US & 36.56$^\circ$ N, 118.77$^\circ$ W & -- & Sequoiadendron giganteum \\
\tcr{\cite{Buntgen2017}, Figure 1} & \tcr{IS} & \tcr{63.71$^\circ$ N,	20.12$^\circ$ W} & \tcr{--} & \tcr{Betula pubescens L.} \\
\cite{Buntgen2018} & PAT02, \tcr{CL} & 32.\tcr{65}$^\circ$ S, 70.\tcr{82}$^\circ$ W & 1980 & Austrocedrus chilensis\\
\textit{Data availability} section & PAT03, \tcr{CL} & 41.\tcr{90}$^\circ$ S, 72.\tcr{68}$^\circ$ W & 783 & Fitzroya cupressoides \\
 & DAR01, NZ & 35.\tcr{92}$^\circ$ S, 173.\tcr{80}$^\circ$ E & 40 & Agathis australis \\
 & \tcr{DAR02, NZ} & \tcr{35.92$^\circ$ S, 173.80$^\circ$ E} & \tcr{40} & \tcr{Agathis australis} \\
 & NEW01, NZ  & 42.\tcr{5}0$^\circ$ S, 171.\tcr{63}$^\circ$ W & 110 & Lagarostrobos colensoi \\
 & TAS01, AU & 41.\tcr{7}5$^\circ$ S, 145.\tcr{2}3$^\circ$ E & 150 & Lagarostrobos franklinii \\
 & CHN01, CN & 37.\tcr{45}$^\circ$ N, 97.\tcr{67}$^\circ$ E & 4019 & Juniperus przewalskii \\
 & JAP01, JP & 30.\tcr{33}$^\circ$ N, 130.\tcr{50}$^\circ$ E & 885 & Cryptomeria japonica \\
 & PAK04, PK & 35.1\tcr{7}$^\circ$ N, 75.\tcr{5}0$^\circ$ E & 3700 & Juniperus excelsa \\
 & TIB01, CN & 31.\tcr{12}$^\circ$ N, 97.\tcr{13}$^\circ$ E & 4000 & Juniperus tibetica \\
 & USA02, US & 35.\tcr{95}$^\circ$ N, 108.\tcr{11}$^\circ$ W & 2357 & Pseudotsuga menziesii \\
 & USA18, US & 37.77$^\circ$ N, 118.\tcr{77}$^\circ$ W & 3539 & Pinus longaeva \\
 & ALB01, AL & 41.\tcr{80}$^\circ$ N, 20.\tcr{23}$^\circ$ E & 1900 & Pinus heldreichii \\
 & AUT02, AT & 46.\tcr{95}$^\circ$ N, 12.\tcr{10}$^\circ$ E & 2170 & Pinus cembra \\
 & CAN06, CA & 54.\tcr{47}$^\circ$ N, 70.\tcr{40}$^\circ$ W & 537 & Picea mariana \\
 & GER07, DE & 48.\tcr{80}$^\circ$ N, 12.\tcr{90}$^\circ$ W & 320 & Abies alba  \\
 & GRE02, GR & 40.\tcr{11}$^\circ$ N, 20.\tcr{97}$^\circ$ E & 2146 & Pinus heldreichii  \\
 & ITA09, IT & 46.\tcr{30}$^\circ$ N, 11.\tcr{03}$^\circ$ E & 1805 & Picea abies \\
 & MON03, MN & 46.\tcr{67}$^\circ$ N, 101.\tcr{77}$^\circ$ E & 2100 & Larix sibirica  \\
 & MON09, MN & 48.\tcr{17}$^\circ$ N, 99.\tcr{87}$^\circ$ E & 2064 & Pinus sibirica  \\
 & ALT01, RU & 50.\tcr{30}$^\circ$ N, 90.\tcr{30}$^\circ$ E & 2300 & Larix sibirica  \\
 & SUI01, CH  & 46.\tcr{62}, 10.\tcr{42}$^\circ$ E & 1250 & Larix decidua  \\
 & USA11, US & 58.\tcr{63}$^\circ$ N, 136.\tcr{47}$^\circ$ W & 425 & Tsuga mertensiana  \\
 & RUS04, RU & 67.\tcr{29}$^\circ$ N, 70.\tcr{42}$^\circ$ E & 35 & Larix sibirica  \\
 & RUS15, RU & 72.\tcr{22}$^\circ$ N, 102.\tcr{63}$^\circ$ E & 250 & Larix gmelinii  \\
 & RUS17, RU & 69.\tcr{00}$^\circ$ N, 148.\tcr{00}$^\circ$ E & 150 & Larix cajanderi  \\
 & RUS20, RU & 67.\tcr{52}$^\circ$ N, 70.\tcr{67}$^\circ$ E & 350 & Larix sibirica \\
 & SWE01, SE & 68.\tcr{27}$^\circ$ N, 19.\tcr{45}$^\circ$ E & 425 & Pinus sylvestris  \\
 & SWE05, SE & 63.\tcr{16}$^\circ$ N, 13.\tcr{57}$^\circ$ E & 530 & Picea abies \\
 & USA10, US & 61.\tcr{12}$^\circ$ N, 147.\tcr{05}$^\circ$ W & 203 & Tsuga mertensiana  \\
\tcr{\cite{Uusitalo2018}, Tab. S3} & \tcr{FI$^\dagger$} & \tcr{46.68$^\circ$ N,	101.77$^\circ$ E} & \tcr{0} & \tcr{Pinus sylvestris} \\
\cite{Scifo2019}, Tab. 1 & UK & 51.\tcr{97}$^\circ$ N, \tcr{5.33}$^\circ$ E & -- & Quercus robur \\
 & US & 38.00$^\circ$ N, 119.\tcr{5}0$^\circ$ W &  -- & Juniperus sp.\\
\tcr{\cite{Pearl2020}, Fig. 6}& \tcr{US} & \tcr{41.76$^\circ$ N,	71.31$^\circ$ W}& \tcr{--}&\tcr{Chamaecyparis thyoides}\\
\tcr{\cite{Hakozaki2020}, pers. comm.}& \tcr{JP} & \tcr{41.16$^\circ$ N,	141.24$^\circ$ E}&\tcr{--}&\tcr{Thujopsis}\\
\tcr{\cite{Walker2023}, Table 2} &\tcr{AU}&\tcr{43.48$^\circ$ S,	146.76$^\circ$ E}&\tcr{--} &\tcr{Athrotaxis selaginoides}\\
\cite{walker2025}, Tab. S2 & MN\tcr{$^*$} & 48.10$^\circ$ N, 99.22$^\circ$ E & -- & Pinus Sibirica \\
 & MN\tcr{$^\dagger$} & 46.40$^\circ$ N, 101.46$^\circ$ W &  -- & Larix Sibirica  \\
 \hline
\end{tabular}
\end{table}

\begin{table}
\caption*{Table~\ref{tab:meta} continued.}
\footnotesize
\begin{tabular}{|l|r|cc|c|}
\hline
Reference & Sample & Coordinates &Altitude, m& Tree \tcr{species} \\
\hline
\multicolumn{5}{|c|}{AD 993 (957 BP)}\\
\hline
\cite{Miyake2013}, \tcr{Table S2}& JP & 30.\tcr{20}$^\circ$ N, 130.\tcr{30}$^\circ$ E & -- & \tcr{Cryptomeria} \\
\cite{Fogtmann-Schulz}, SI & Mojbol, DK\tcr{$^\dagger$} & 55.\tcr{30}$^\circ$ N, 9.1\tcr{7}$^\circ$ E & -- & Quercus\\
\cite{Buntgen2018} & PAT03, CL & 41.90$^\circ$ S, 72.68$^\circ$ W & 783 & Fitzroya cupressoides\\
\textit{Data availability} section & DAR01, NZ & 35.92$^\circ$ S, 173.80$^\circ$ E & 40 & Agathis australis \\
& CHI01, CN & 37.45$^\circ$ N, 97.67$^\circ$ E & 4019 & Juniperus przewalskii \\
 & MON05, MN & 46.67$^\circ$ N, 101.77$^\circ$ E & 2100 & Larix sibirica \\
 & ALT02, RU & 50.\tcr{3}0$^\circ$ N, 90.\tcr{3}0$^\circ$ E & 2300 & Larix sibirica \\
 & SWE01, SE & 68.27$^\circ$ N, 19.45$^\circ$ E & 425 & Pinus sylvestris \\
 & USA07, US & 35.96$^\circ$ N, 108.11$^\circ$ W & 2358 & Pseudotsuga menziesii  \\
 \cite{Scifo2019}, Tab. 1 & US & 38.00$^\circ$ N, 119.\tcr{5}0$^\circ$ W &  -- & Juniperus sp.\\
\tcr{\cite{Rakowski2018}}, Tab.1 & Grabie, PL & 50.0\tcr{4}$^\circ$ N, 19.\tcr{99}$^\circ$ E & -- & Quercus robur L.\\
\cite{Hakozaki2020}, SI & JP & 41.9$^\circ$ N, 141.14$^\circ$ E & -- & Thujopsis \\
\tcr{\cite{Brehm2021}, SM Fig. 1}& \tcr{UK} & \tcr{51.75$^\circ$ N,	0.34$^\circ$ W}& \tcr{--}& \tcr{Quercus sp.}\\
\tcr{\cite{Miyake2022}, SM} & \tcr{JP} & \tcr{30.20$^\circ$ N,	130.30$^\circ$ E} & \tcr{--} & \tcr{Cryptomeria} \\
\cite{walker2025}, Tab. S2&  MN\tcr{$^*$} & 48.1\tcr{7}$^\circ$ N, 99.\tcr{87}$^\circ$ E & -- & Pinus Sibirica \\
& \tcr{MN$^*$} & \tcr{46.68$^\circ$ N, 101.77$^\circ$ E} & -- & \tcr{Pinus Sibirica} \\
& \tcr{MN$^\dagger$} & \tcr{48.17$^\circ$ N, 99.87$^\circ$ E} & \tcr{--} & \tcr{Larix Sibirica} \\
\hline
\end{tabular}
\end{table}

\newpage

\section*{Appendix B. Table of response functions}
\label{S:AppendixB}

Response functions of $\Delta^{14}$C (relative radiocarbon concentration in near-surface air, in per mil) for a reference extreme solar particle event (ESPE) under Holocene pre-industrial conditions, computed using the SOCOL:14C-Ex model, are available on Zenodo via the doi: doi.org/10.5281/zenodo.17397487.

\newpage
\printbibliography

@article{ADAMS19983,
title = {A new estimate of changing carbon storage on land since the last glacial maximum, based on global land ecosystem reconstruction},
journal = {Global and Planetary Change},
volume = {16-17},
pages = {3-24},
year = {1998},
issn = {0921-8181},
%doi = {https://doi.org/10.1016/S0921-8181(98)00003-4},
url = {https://www.sciencedirect.com/science/article/pii/S0921818198000034},
author = {J.M. Adams and H. Faure}}

@article{Hughen1998,
  author  = {Hughen, K. A. and Overpeck, J. T. and Lehman, S. J. and Kashgarian, M. and Southon, J. and Peterson, L. C. and Alley, R. and Sigman, D. M.},
  title   = {Deglacial changes in ocean circulation from an extended radiocarbon calibration},
  journal = {Nature},
  year    = {1998},
  volume  = {391},
  pages   = {65--68},
  %doi     = {10.1038/34150}
}

@article{Goslar1995,
  author  = {Goslar, T. and others},
  title   = {High concentration of atmospheric {$^{14}$C} during the Younger Dryas cold episode},
  journal = {Nature},
  year    = {1995},
  volume  = {377},
  pages   = {414--417},
  %doi     = {10.1038/377414a0}
}

@article{Rakowski2018,
author = {Rakowski, Andrzej Z. and Kraepiec, Marek and Huels, Mathias and Pawlyta, Jacek and Boudin, Mathieu},
%doi = {10.1017/RDC.2018.74},
file = {:Users/sergeykoldobskiy/Library/Application Support/Mendeley Desktop/Downloaded/Rakowski et al. - 2018 - Increase in radiocarbon concentration in tree rings from kujawy village (Se Poland) Around Ad 993-994.pdf:pdf},
issn = {00338222},
journal = {Radiocarbon},
keywords = {AMS,Miyake effect,atmosphere,biosphere,calibration curve,cosmogenic isotopes,radiocarbon,tree rings},
number = {4},
pages = {1249--1258},
title = {{Increase in radiocarbon concentration in tree rings from kujawy village (Se Poland) Around Ad 993-994}},
volume = {60},
year = {2018}
}

@article{Miyake2022,
author = {Miyake, Fusa and Hakozaki, Masataka and Kimura, Katsuhiko and Tokanai, Fuyuki and Nakamura, Toshio and Takeyama, Mirei and Moriya, Toru},
%doi = {10.3389/fspas.2022.886140},
file = {:Users/sergeykoldobskiy/Library/Application Support/Mendeley Desktop/Downloaded/Miyake et al. - 2022 - Regional Differences in Carbon-14 Data of the 993 CE Cosmic Ray Event.pdf:pdf},
issn = {2296-987X},
journal = {Front. Astron. Sp. Sci.},
keywords = {SEP event,cosmogenic nuclides,dendrochronology,radiocarbon dating,solar activity,tree ring},
number = {July},
pages = {1--8},
title = {{Regional Differences in Carbon-14 Data of the 993 CE Cosmic Ray Event}},
url = {https://www.frontiersin.org/articles/10.3389/fspas.2022.886140/full},
volume = {9},
year = {2022}
}

@article{Rakowski2024,
author = {Rakowski, Andrzej Z. and Pawlyta, Jacek and Miyahara, Hiroko and Kraepiec, Marek and Moln{\'{a}}r, Mih{\'{a}}ly and Wiktorowski, Damian and Minami, Masayo},
%doi = {10.1017/RDC.2023.79},
file = {:Users/sergeykoldobskiy/Library/Application Support/Mendeley Desktop/Downloaded/Rakowski et al. - 2024 - Radiocarbon concentration in sub-annual tree rings from Poland around 660 BCE.pdf:pdf;:Users/sergeykoldobskiy/Library/Application Support/Mendeley Desktop/Downloaded/Rakowski et al. - 2024 - Radiocarbon concentration in sub-annual tree rings from Poland around 660 BCE.docx:docx},
issn = {00338222},
journal = {Radiocarbon},
keywords = {Miyake Event,SEP,calibration curves,radiocarbon dating,tree rings},
number = {6},
pages = {1981--1990},
title = {{Radiocarbon concentration in sub-annual tree rings from Poland around 660 BCE}},
volume = {66},
year = {2024}
}

@article{DELAYGUE2003,
title = {Simulation of atmospheric radiocarbon during abrupt oceanic circulation changes: trying to reconcile models and reconstructions},
journal = {Quaternary Science Reviews},
volume = {22},
number = {15},
pages = {1647-1658},
year = {2003},
issn = {0277-3791},
%doi = {https://doi.org/10.1016/S0277-3791(03)00171-9},
url = {https://www.sciencedirect.com/science/article/pii/S0277379103001719},
author = {Gilles Delaygue and Thomas F Stocker and Fortunat Joos and Gian-Kasper Plattner}}

@article{Singarayer2008,
author = {Singarayer, Joy S. and Richards, David A. and Ridgwell, Andy and Valdes, Paul J. and Austin, William E. N. and Beck, J. Warren},
title = {An oceanic origin for the increase of atmospheric radiocarbon during the Younger Dryas},
journal = {Geophysical Research Letters},
volume = {35},
number = {14},
pages = {},
keywords = {Earth system model, radiocarbon, Younger Dryas},
%doi = {https://doi.org/10.1029/2008GL034074},
url = {https://agupubs.onlinelibrary.wiley.com/doi/abs/10.1029/2008GL034074},
eprint = {https://agupubs.onlinelibrary.wiley.com/doi/pdf/10.1029/2008GL034074},
year = {2008}
}

@article{Pearl2020,
author = {Pearl, Jessie K. and Anchukaitis, Kevin J. and Donnelly, Jeffrey P. and Pearson, Charlotte and Pederson, Neil and {Lardie Gaylord}, Mary C. and McNichol, Ann P. and Cook, Edward R. and Zimmermann, George L.},
%doi = {10.1016/j.quascirev.2019.106104},
file = {:Users/sergeykoldobskiy/Library/Application Support/Mendeley Desktop/Downloaded/Pearl et al. - 2020 - A late Holocene subfossil Atlantic white cedar tree-ring chronology from the northeastern United States.pdf:pdf},
issn = {02773791},
journal = {Quat. Sci. Rev.},
keywords = {Coastal,Geomorphology,Holocene,North America,Paleoclimatology,Radiogenic isotopes,Tree-rings},
pages = {106104},
publisher = {Elsevier Ltd},
title = {{A late Holocene subfossil Atlantic white cedar tree-ring chronology from the northeastern United States}},
url = {https://doi.org/10.1016/j.quascirev.2019.106104},
volume = {228},
year = {2020}
}

@article{Rakowski2015,
author = {Rakowski, Andrzej Z. and Pawlyta, Jacek and Miyahara, Hiroko and Kra\k{e}piec, Marek and Moln{\'{a}}r, Mih{\'{a}}ly and Wiktorowski, Damian and Minami, Masayo},
%doi = {10.1016/j.nimb.2015.03.035},
file = {:Users/sergeykoldobskiy/Library/Application Support/Mendeley Desktop/Downloaded/Rakowski et al. - 2015 - Increase of radiocarbon concentration in tree rings from Kujawy (SE Poland) around AD 774-775.pdf:pdf},
issn = {0168583X},
journal = {Nucl. Instruments Methods Phys. Res. Sect. B Beam Interact. with Mater. Atoms},
keywords = {Atmosphere,Calibration curve,Cosmogenic isotopes,Radiocarbon,Tree rings},
pages = {564--568},
title = {{Increase of radiocarbon concentration in tree rings from Kujawy (SE Poland) around AD 774-775}},
volume = {361},
year = {2015}
}

@article{Walker2023,
author = {Walker, Meagan and Mueller, Alexis and Allen, Kathy and Fenwick, Pavla and Agrawal, Vikas and Anchukaitis, Kevin and Hessl, Amy},
%doi = {10.1016/j.dendro.2022.126048},
file = {:Users/sergeykoldobskiy/Library/Application Support/Mendeley Desktop/Downloaded/Walker et al. - 2023 - High resolution radiocarbon spike confirms tree ring dating with low sample depth.pdf:pdf},
issn = {16120051},
journal = {Dendrochronologia},
keywords = {Australia,Earlywood and latewood,King Billy pine,Subannual tree ring,Tasmania,$\Delta$14C in 775 CE},
number = {October 2022},
pages = {126048},
publisher = {Elsevier GmbH},
title = {{High resolution radiocarbon spike confirms tree ring dating with low sample depth}},
url = {https://doi.org/10.1016/j.dendro.2022.126048},
volume = {77},
year = {2023}
}

@article{Buntgen2017,
author = {B{\"{u}}ntgen, Ulf and Eggertsson, {\'{O}}lafur and Wacker, Lukas and Sigl, Michael and Ljungqvist, Fredrik Charpentier and di Cosmo, Nicola and Plunkett, Gill and Krusic, Paul J. and Newfield, Timothy P. and Esper, Jan and Lane, Christine and Reinig, Frederick and Oppenheimer, Clive},
%doi = {10.1130/G39269.1},
file = {:Users/sergeykoldobskiy/Library/Application Support/Mendeley Desktop/Downloaded/B{\"{u}}ntgen et al. - 2017 - Multi-proxy dating of Iceland's major pre-settlement Katla eruption to 822-823 CE.pdf:pdf},
issn = {19432682},
journal = {Geology},
number = {9},
pages = {783--786},
title = {{Multi-proxy dating of Iceland's major pre-settlement Katla eruption to 822-823 CE}},
url = {http://pubs.geoscienceworld.org/geology/article/45/9/783/208004/Multiproxy-dating-of-Icelands-major-presettlement},
volume = {45},
year = {2017}
}

@article{Miller2025,
author = {Miller, John B. and Lehman, Scott J. and Lindsay, Colin M.},
title = {Numerical Representation of Contemporary Atmospheric 14CO2: 1. Time-Varying Global Fluxes and Atmospheric Mass Balance},
journal = {Global Biogeochemical Cycles},
volume = {39},
number = {11},
pages = {e2025GB008522},
keywords = {radiocarbon, CO2, atmospheric mass balance},
%doi = {https://doi.org/10.1029/2025GB008522},
url = {https://agupubs.onlinelibrary.wiley.com/doi/abs/10.1029/2025GB008522},
year = {2025}}

@article{Masarik2009,
author = {Masarik, J. and Beer, J.},
title = {An updated simulation of particle fluxes and cosmogenic nuclide production in the Earth's atmosphere},
journal = {Journal of Geophysical Research: Atmospheres},
volume = {114},
number = {D11},
pages = {},
keywords = {simulation, particle fluxes},
%doi = {https://doi.org/10.1029/2008JD010557},
url = {https://agupubs.onlinelibrary.wiley.com/doi/abs/10.1029/2008JD010557},
year = {2009}}

@ARTICLE{usoskin_AA_25,
       author = {{Usoskin}, I. and {Chatzistergos}, T. and {Solanki}, S.K. and {Krivova}, N. and {Kovaltsov}, G. and {Brehm}, N. and {Christl}, M. and {Wacker}, L.},
        title = "{Sunspot cycles for the first millennium BC reconstructed from radiocarbon}",
      journal = {Astron. Astrophys.},
     keywords = {Sun: activity, Sun: magnetic fields, sunspots},
         year = 2025,
       volume = {698},
          eid = {A182},
        pages = "{A182}",
          %doi = {10.1051/0004-6361/202555148},
       adsurl = {https://ui.adsabs.harvard.edu/abs/2025A&A...698A.182U},
      adsnote = {Provided by the SAO/NASA Astrophysics Data System}
}

@article{Bard1997,
  author    = {E. Bard and G. M. Raisbeck and F. Yiou and J. Jouzel},
  title     = {Solar modulation of cosmogenic nuclide production over the last millennium: comparison between \textsuperscript{14}C and \textsuperscript{10}Be records},
  journal   = {Earth and Planetary Science Letters},
  year      = {1997},
  volume    = {150},
  pages     = {453--462},
  %doi       = {10.1016/S0012-821X(97)00082-4}
}

@article{Fahrni2020,
abstract = {As part of the ongoing effort to improve the Northern Hemisphere radiocarbon (C) calibration curve, this study investigates the period of 856 BC to 626 BC (2805-2575 yr BP) with a total of 403 single-year C measurements. In this age range, IntCal13 was constructed largely from German and Irish oak as well as Californian bristlecone pine C dates, with most samples measured with a 10-yr resolution. The new data presented here is the first atmospheric C single-year record of the older end of the Hallstatt plateau based on an absolutely dated tree-ring chronology. The data helped reveal a major solar proton event (SPE) which caused a spike in the production rate of cosmogenic radionuclides around 2610/2609 BP. This production event is thought to have reached a magnitude similar to the 774/775 AD production event but has remained undetected due to averaging effects in the decadal calibration data. The record leading up to the 2610/2609 BP event reveals a 11-yr solar cycle with varying cyclicity. Features of the new data and the benefits of higher resolution calibration are discussed.},
author = {Fahrni, Simon M. and Southon, John and Fuller, Benjamin T. and Park, Junghun and Friedrich, Michael and Muscheler, Raimund and Wacker, Lukas and Taylor, R. E.},
%doi = {10.1017/RDC.2020.16},
file = {:Users/sergeykoldobskiy/Downloads/single-year-german-oak-and-californian-bristlecone-pine-14c-data-at-the-beginning-of-the-hallstatt-plateau-from-856-bc-to-626-bc.pdf:pdf},
issn = {00338222},
journal = {Radiocarbon},
keywords = {IntCal,calibration,radiocarbon,tree rings},
number = {4},
pages = {919--937},
title = {{Single-Year German oak and Californian Bristlecone Pine C Data at the Beginning of the Hallstatt Plateau from 856 BC to 626 BC}},
volume = {62},
year = {2020}
}

@Article{Jeltsch2019,
AUTHOR = {Jeltsch-Th\"ommes, A. and Battaglia, G. and Cartapanis, O. and Jaccard, S. L. and Joos, F.},
TITLE = {Low terrestrial carbon storage at the Last Glacial Maximum: constraints from multi-proxy data},
JOURNAL = {Climate of the Past},
VOLUME = {15},
YEAR = {2019},
NUMBER = {2},
PAGES = {849--879},
URL = {https://cp.copernicus.org/articles/15/849/2019/},
%DOI = {10.5194/cp-15-849-2019}
}

@article{Prentice1990,
  author    = {Prentice, K. C. and Fung, I. Y.},
  title     = {The sensitivity of terrestrial carbon storage to climate change},
  journal   = {Nature},
  volume    = {346},
  pages     = {48--51},
  year      = {1990},
  %doi       = {10.1038/346048a0},
  url       = {https://doi.org/10.1038/346048a0}
}

@ARTICLE{bieber2013,
       author = {{Bieber}, J.W. and {Clem}, J. and {Evenson}, P. and {Pyle}, R. and {S{\'a}iz}, A. and {Ruffolo}, D.},
        title = "{Giant Ground Level Enhancement of Relativistic Solar Protons on 2005 January 20. I. Spaceship Earth Observations}",
      journal = {Astrophys. J.},
     keywords = {interplanetary medium, solar-terrestrial relations, Sun: coronal mass ejections: CMEs, Sun: particle emission},
         year = 2013,
       volume = {771},
       number = {2},
          eid = {92},
        pages = {92},
          %doi = {10.1088/0004-637X/771/2/92},
       adsurl = {https://ui.adsabs.harvard.edu/abs/2013ApJ...771...92B},
      adsnote = {Provided by the SAO/NASA Astrophysics Data System}
}

@article{Usoskin2023a,
    title = {{A history of solar activity over millennia}},
    year = {2023},
    journal = {Living Reviews in Solar Physics},
    author = {Usoskin, Ilya G.},
    number = {1},
    month = {5},
    pages = {2},
    volume = {20},
    publisher = {Springer International Publishing},
    url = {https://doi.org/10.1007/s41116-023-00036-z https://link.springer.com/10.1007/s41116-023-00036-z},
    isbn = {0123456789},
    %doi = {10.1007/s41116-023-00036-z},
    issn = {1614-4961},
    keywords = {Cosmogenic isotopes, Long-term reconstructions, Paleo-astrophysics, Solar activity, Solar dynamo, Solar physics, Solarterrestrial relations}
}

@article{Bard2023,
    title = {{A radiocarbon spike at 14 300 cal yr BP in subfossil trees provides the impulse response function of the global carbon cycle during the Late Glacial}},
    year = {2023},
    journal = {Philosophical Transactions of the Royal Society A: Mathematical, Physical and Engineering Sciences},
    author = {Bard, Edouard and Miramont, Cécile and Capano, Manuela and Guibal, Frédéric and Marschal, Christian and Rostek, Frauke and Tuna, Thibaut and Fagault, Yoann and Heaton, Timothy J},
    number = {2261},
    month = {11},
    pages = {20220206},
    volume = {381},
    url = {https://royalsocietypublishing.org/doi/10.1098/rsta.2022.0206},
    isbn = {0000000272},
    %doi = {10.1098/rsta.2022.0206},
    issn = {1364-503X},
    keywords = {climatology, geochemistry}
}

@ARTICLE{miyake12,
   author = {Miyake, F. and Nagaya, K. and Masuda, K. and Nakamura, T.},
    title = "{A signature of cosmic-ray increase in ad 774-775 from tree rings in Japan}",
  journal = {Nature},
     year = 2012,
   volume = {486},
    pages = {240-242},
      %doi = {10.1038/nature11123},
   adsurl = {http://adsabs.harvard.edu/abs/2011ACPD...1114003M},
  adsnote = {Provided by the SAO/NASA Astrophysics Data System}
}

@article{Miyake2021,
    title = {{A Single‐Year Cosmic Ray Event at 5410 BCE Registered in 14C of Tree Rings}},
    year = {2021},
    journal = {Geophysical Research Letters},
    author = {Miyake, F. and Panyushkina, I. P. and Jull, A. J. T. and Adolphi, F. and Brehm, N. and Helama, S. and Kanzawa, K. and Moriya, T. and Muscheler, R. and Nicolussi, K. and Oinonen, M. and Salzer, M. and Takeyama, M. and Tokanai, F. and Wacker, L.},
    number = {11},
    month = {6},
    pages = {e2021GL093419},
    volume = {48},
    url = {https://onlinelibrary.wiley.com/doi/10.1029/2021GL093419},
    %doi = {10.1029/2021GL093419},
    issn = {0094-8276},
    keywords = {cosmogenic nuclide, radiocarbon, solar activity, solar energetic particle, tree rings}
}

@article{reimer20,
Author = {Reimer, P.J. and Austin, W.E.N. and Bard, E. and
   Bayliss, A. and Blackwell, P.G. and Ramsey, C.B. and
   Butzin, M. and Cheng, H. and Edwards, R.L. and Friedrich, M. and
   Grootes, P.M. and Guilderson, T.P. and Hajdas, I. and Heaton, T.J. and
   Hogg, A.G. and Hughen, K.A. and Kromer, B. and Manning, S.W. and Muscheler, R.
   and Palmer, J.G. and Pearson, C. and van der Plicht, J. and Reimer, R.W. and
   Richards, D.A. and Scott, E.M. and Southon, J.R. and Turney, C.S.M. and
   Wacker, L. and Adolphi, F. and Buentgen, U. and Capano, M. and Fahrni, S.M. and
   Fogtmann-Schulz, A. and Friedrich, R. and Koehler, P. and Kudsk, S. and
   Miyake, F. and Olsen, J. and Reinig, F. and Sakamoto, M. and Sookdeo, A. and Talamo, S.},
Title = {{The INTCAL20 Northern hemisphere radiocarbon age calibration curve (0-55 CAL KBP)}},
Journal = {{Radiocarbon}},
year = {2020},
Volume = {{62}},
Number = {{4}},
Pages = {{725-757}},
%DOI = {{10.1017/RDC.2020.41}}
}

@article{Raukunen2022,
    title = {{Annual integral solar proton fluences for 1984–2019}},
    year = {2022},
    journal = {Astronomy {\&} Astrophysics},
    author = {Raukunen, O. and Usoskin, I. and Koldobskiy, S. and Kovaltsov, G. and Vainio, R.},
    month = {9},
    pages = {A65},
    volume = {665},
    url = {https://www.aanda.org/10.1051/0004-6361/202243736},
    %doi = {10.1051/0004-6361/202243736},
    issn = {0004-6361},
    keywords = {activity, flares, particle emission, solar-terrestrial relations, sun}
}

@article{Miyake2013,
    title = {{Another rapid event in the carbon-14 content of tree rings}},
    year = {2013},
    journal = {Nature Communications},
    author = {Miyake, F. and Masuda, Kimiaki and Nakamura, Toshio},
    number = {1},
    month = {4},
    pages = {1748},
    volume = {4},
    url = {https://www.nature.com/articles/ncomms2783},
    %doi = {10.1038/ncomms2783},
    issn = {2041-1723}
}

@article{Sukhodolov2017,
    title = {{Atmospheric impacts of the strongest known solar particle storm of 775 AD}},
    year = {2017},
    journal = {Scientific Reports},
    author = {Sukhodolov, Timofei and Usoskin, Ilya and Rozanov, Eugene and Asvestari, Eleanna and Ball, William T. and Curran, Mark A.J. and Fischer, Hubertus and Kovaltsov, Gennady and Miyake, Fusa and Peter, Thomas and Plummer, Christopher and Schmutz, Werner and Severi, Mirko and Traversi, Rita},
    pages = {1--9},
    volume = {7},
    publisher = {Nature Publishing Group},
    %doi = {10.1038/srep45257},
    issn = {20452322},
    pmid = {28349934}
}

@article{Fogtmann-Schulz,
author = {Fogtmann-Schulz, A. and Østbø, S. M. and Nielsen, S. G. B. and Olsen, J. and Karoff, C. and Knudsen, M. F.},
title = {Cosmic ray event in 994 C.E. recorded in radiocarbon from Danish oak},
journal = {Geophysical Research Letters},
volume = {44},
number = {16},
pages = {8621-8628},
keywords = {radiocarbon, tree rings, Miyake event, astrochronology, solar proton event, space environment},
%doi = {https://doi.org/10.1002/2017GL074208},
year = {2017}}

@inproceedings{Hakozaki2020,
  author    = {Hakozaki, M. and Miyake, F. and Nakamura, T.},
  title     = {775 and 994 \({}^{14}\)C Events in the Tree-Rings of Northern Japanese Trees},
  booktitle = {Proceedings of EA-AMS 8 \& JAMS22},
  month     = jun,
  year      = {2020},
  pages     = {89--90}
}

@article{weisenstein1997sulfur,
  author = {Weisenstein, D. K. and Yue, G. K. and Ko, M. K. W. and Sze, N.-D. and Rodriguez, J. M. and Scott, C. J.},
  title = "{A two-dimensional model of sulfur species and aerosols}",
  journal = {Journal of Geophysical Research: Atmospheres},
  volume = {102},
  number = {D11},
  pages = {13019--13035},
  year = {1997},
  %doi = {10.1029/97JD00901}
}

@article{egorova2003mezon,
  author = {Egorova, T. and Rozanov, E. and Zubov, V. and Karol, I.},
  title = "{Model for investigating ozone trends MEZON}",
  journal = {Izvestiya – Atmospheric and Oceanic Physics},
  volume = {39},
  pages = {277--292},
  year = {2003}
}

@article{hommel2011maecham5,
  author = {Hommel, R. and Timmreck, C. and Graf, H. F.},
  title = "{The global middle-atmosphere aerosol model MAECHAM5-SAM2: comparison with satellite and in-situ observations}",
  journal = {Geoscientific Model Development},
  volume = {4},
  number = {3},
  pages = {809--834},
  year = {2011},
  %doi = {10.5194/gmd-4-809-2011}
}

@article{marcott2014carbon,
  author = {Marcott, S. and Bauska, T. and Buizert, C. and Steig, E. and Rosen, J. and Cuffey, K. and Brook, E. J.},
  title = "{Centennial-scale changes in the global carbon cycle during the last deglaciation}",
  journal = {Nature},
  volume = {514},
  number = {7524},
  pages = {616--619},
  year = {2014},
  %doi = {10.1038/nature13799}
}

@article{Indermuehle1999,
  author    = {Inderm{\"u}hle, A. and Stocker, T. F. and Joos, F. and Fischer, H. and Smith, H. J. and Wahlen, M. and Deck, B. and Mastroianni, D. and Tschumi, J. and Blunier, T. and Meyer, R. and Stauffer, B.},
  title     = {Holocene carbon-cycle dynamics based on CO$_2$ trapped in ice at Taylor Dome, Antarctica},
  journal   = {Nature},
  volume    = {398},
  pages     = {121--126},
  year      = {1999},
  publisher = {Nature Publishing Group},
  %doi       = {10.1038/18158},
  url       = {https://www.nature.com/articles/18158}
}

@ARTICLE{stuiver77,
   author = {{Stuiver}, M. and {Pollach}, H.},
    title = "{Discussion: Reporting of $^{14}$C data}",
  journal = {Radiocarbon},
     year = 1977,
   volume = 19,
    pages = {355--363}
}

@article{sheng2015sulfur,
  author = {Sheng, J.-X. and Weisenstein, D. K. and Luo, B.-P. and Rozanov, E. and Stenke, A. and Anet, J. and Peter, T.},
  title = "{Global atmospheric sulfur budget under volcanically quiescent conditions: Aerosol-chemistry-climate model predictions and validation}",
  journal = {Journal of Geophysical Research: Atmospheres},
  volume = {120},
  number = {1},
  pages = {256--276},
  year = {2015},
  %doi = {10.1002/2014JD021985}
}

@article{walker2025,
author = {Walker, M. and Shobe, C. and Andrea‐Hayles, L. and Dey, L. and Suran, Byambagerel and Baatarbileg, Nachin and Hessl, A.},
year = {2025},
month = {06},
pages = {},
title = {Reconstructing Annual $\Delta$C During Miyake Events Using Deciduous and Evergreen Trees},
volume = {39},
journal = {Global Biogeochemical Cycles},
%doi = {10.1029/2024GB008423}
}

@article{Scifo2019,
  author       = {Scifo, A. and Kuitems, M. and Neocleous, A. and et al.},
  title        = {Radiocarbon Production Events and their Potential Relationship with the Schwabe Cycle},
  journal      = {Scientific Reports},
  volume       = {9},
  pages        = {17056},
  year         = {2019},
  %doi          = {10.1038/s41598-019-53296-x},
  url          = {https://doi.org/10.1038/s41598-019-53296-x}
}

@article{Rakowski_Krąpiec_Huels_Pawlyta_Hamann_Wiktorowski_2019, title={Abrupt Increase of Radiocarbon Concentration in 660 BC in Tree Rings from Grabie Near Kraków (SE Poland)}, volume={61}, DOI={10.1017/RDC.2019.40}, number={5}, journal={Radiocarbon}, author={Rakowski, Andrzej and Krąpiec, Marek and Huels, Matthias and Pawlyta, Jacek and Hamann, Christian and Wiktorowski, Damian}, year={2019}, pages={1327–1335}}

@article{Paleari2022,
    title = {{Cosmogenic radionuclides reveal an extreme solar particle storm near a solar minimum 9125 years BP}},
    year = {2022},
    journal = {Nature Communications},
    author = {Paleari, Chiara I and Mekhaldi, Florian and Adolphi, Florian and Christl, Marcus and Vockenhuber, Christof and Gautschi, Philip and Beer, Jürg and Brehm, Nicolas and Erhardt, Tobias and Synal, Hans-arno and Wacker, Lukas and Wilhelms, Frank and Muscheler, Raimund},
    number = {1},
    month = {12},
    pages = {214},
    volume = {13},
    publisher = {Springer US},
    url = {https://www.nature.com/articles/s41467-021-27891-4},
    %doi = {10.1038/s41467-021-27891-4},
    issn = {2041-1723}
}

@article{Koldobskiy2022,
    title = {{Effective Energy of Cosmogenic Isotope (10 Be, 14C and 36Cl) Production by Solar Energetic Particles and Galactic Cosmic Rays}},
    year = {2022},
    journal = {Journal of Geophysical Research: Space Physics},
    author = {Koldobskiy, Sergey and Usoskin, Ilya and Kovaltsov, Gennady A.},
    number = {1},
    month = {1},
    pages = {e2021JA029919},
    volume = {127},
    url = {https://onlinelibrary.wiley.com/doi/10.1029/2021JA029919},
    %doi = {10.1029/2021JA029919},
    issn = {2169-9380}
}

@article{Panovska2023,
    title = {{Effects of Global Geomagnetic Field Variations Over the Past 100,000 Years on Cosmogenic Radionuclide Production Rates in the Earth's Atmosphere}},
    year = {2023},
    journal = {Journal of Geophysical Research: Space Physics},
    author = {Panovska, Sanja and Poluianov, Stepan and Gao, Jiawei and Korte, Monika and Mishev, Alexander and Shprits, Yuri Y. and Usoskin, Ilya},
    number = {8},
    month = {8},
    volume = {128},
    url = {https://agupubs.onlinelibrary.wiley.com/doi/10.1029/2022JA031158},
    %doi = {10.1029/2022JA031158},
    issn = {2169-9380}
}

@article{Brehm2021,
    title = {{Eleven-year solar cycles over the last millennium revealed by radiocarbon in tree rings}},
    year = {2021},
    journal = {Nature Geoscience},
    author = {Brehm, Nicolas and Bayliss, Alex and Christl, Marcus and Synal, Hans-Arno and Adolphi, Florian and Beer, Jürg and Kromer, Bernd and Muscheler, Raimund and Solanki, Sami K. and Usoskin, Ilya and Bleicher, Niels and Bollhalder, Silvia and Tyers, Cathy and Wacker, Lukas},
    number = {1},
    month = {1},
    pages = {10--15},
    volume = {14},
    url = {http://www.nature.com/articles/s41561-020-00674-0},
    %doi = {10.1038/s41561-020-00674-0},
    issn = {1752-0894}
}

@article{Jull2014,
    title = {{Excursions in the 14 C record at A.D. 774–775 in tree rings from Russia and America}},
    year = {2014},
    journal = {Geophysical Research Letters},
    author = {Jull, A. J. Timothy and Panyushkina, Irina P. and Lange, Todd E. and Kukarskih, Vladimir V. and Myglan, Vladimir S. and Clark, Kelley J. and Salzer, Matthew W. and Burr, George S. and Leavitt, Steven W.},
    number = {8},
    month = {4},
    pages = {3004--3010},
    volume = {41},
    url = {https://agupubs.onlinelibrary.wiley.com/doi/10.1002/2014GL059874},
    %doi = {10.1002/2014GL059874},
    issn = {0094-8276}
}

@article{Cliver2022,
    title = {{Extreme solar events}},
    year = {2022},
    journal = {Living Reviews in Solar Physics},
    author = {Cliver, Edward W and Schrijver, Carolus J. and Shibata, Kazunari and Usoskin, Ilya G},
    number = {1},
    month = {12},
    pages = {2},
    volume = {19},
    publisher = {Springer International Publishing},
    url = {https://link.springer.com/10.1007/s41116-022-00033-8},
    isbn = {0123456789},
    %doi = {10.1007/s41116-022-00033-8},
    issn = {2367-3648},
    keywords = {Sun,Superflare stars,Solar flares,Coronal mass eje, coronal mass ejections {\'{a}}, energetic particle events {\'{a}}, extreme solar, geomagnetic storms {\'{a}} solar, sun {\'{a}} superflare stars, {\'{a}} solar flares {\'{a}}}
}

@article{Usoskin2023b,
    title = {{Extreme Solar Events: Setting up a Paradigm}},
    year = {2023},
    journal = {Space Science Reviews},
    author = {Usoskin, Ilya and Miyake, Fusa and Baroni, Melanie and Brehm, Nicolas and Dalla, Silvia and Hayakawa, Hisashi and Hudson, Hugh and Jull, A J Timothy and Knipp, Delores and Koldobskiy, Sergey and Maehara, Hiroyuki and Mekhaldi, Florian and Notsu, Yuta and Poluianov, Stepan and Rozanov, Eugene and Shapiro, Alexander and Spiegl, Tobias and Sukhodolov, Timofei and Uusitalo, Joonas and Wacker, Lukas},
    number = {8},
    pages = {73},
    volume = {219},
    url = {http://dx.doi.org/10.1007/s11214-023-01018-1 https://link.springer.com/10.1007/s11214-023-01018-1},
    %doi = {10.1007/s11214-023-01018-1},
    issn = {0038-6308},
    keywords = {cosmogenic isotopes, solar activity, solar flares, stellar flares}
}

@article{Heaton2024,
    title = {{Extreme solar storms and the quest for exact dating with radiocarbon}},
    year = {2024},
    journal = {Nature},
    author = {Heaton, T J and Bard, E and Bayliss, A and Blaauw, M and Bronk Ramsey, C and Reimer, P J and Turney, C S M and Usoskin, I},
    number = {8029},
    month = {9},
    pages = {306--317},
    volume = {633},
    url = {https://www.nature.com/articles/s41586-024-07679-4},
    %doi = {10.1038/s41586-024-07679-4},
    issn = {0028-0836}
}

@article{oeschger74,
  author   = {Oeschger, H. and Siegenthaler, U. and Schotterer, U. and Gugelmann, A.},
  format   = {print},
  journal  = {Tellus},
  pages    = {168--192},
  title    = {A box diffusion model to study the carbon dioxide exchange in nature},
  volume   = {27},
  issue = {2},
  year     = {1975}
}

@article{Liu14,
title={Mysterious abrupt carbon-14 increase in coral contributed by a comet},
DOI={10.1038/srep03728},
journal={Sci. Rep.},
volume = {4},
author={Liu, Y. and Zhang, Z. and Peng, Z. and Ling, M. and Shen, Ch. and Liu, W. and Sun, X. and Shen, Ch. and Liu, K. and Sun, W.},
pages="{3728}",
eid={3728},
year={2014}
}

@ARTICLE{pavlov13,
   author = {{Pavlov}, A.K. and {Blinov}, A.V. and {Konstantinov}, A.N. and
	{Ostryakov}, V.M. and {Vasilyev}, G.I. and {Vdovina}, M.A. and
	{Volkov}, P.A.},
    title = "{AD 775 pulse of cosmogenic radionuclides production as imprint of a Galactic gamma-ray burst}",
  journal = {Mon. Notes R. Astron. Soc.},
 keywords = {Earth, gamma-ray burst: general, Galaxy: general},
     year = 2013,
   volume = 435,
    pages = {2878-2884},
      %doi = {10.1093/mnras/stt1468},
   adsurl = {http://adsabs.harvard.edu/abs/2013MNRAS.435.2878P},
  adsnote = {Provided by the SAO/NASA Astrophysics Data System}
}

@article{Verger2023,
    title = {{GEOV2: Improved smoothed and gap filled time series of LAI, FAPAR and FCover 1 km Copernicus Global Land products}},
    year = {2023},
    journal = {International Journal of Applied Earth Observation and Geoinformation},
    author = {Verger, Aleixandre and S{\'{a}}nchez-Zapero, Jorge and Weiss, Marie and Descals, Adrià and Camacho, Fernando and Lacaze, Roselyne and Baret, Frédéric},
    pages = {103479},
    volume = {123},
    url = {https://www.sciencedirect.com/science/article/pii/S1569843223003035},
    %doi = {https://doi.org/10.1016/j.jag.2023.103479},
    issn = {1569-8432},
    keywords = {Fraction of absorbed PAR, Global vegetation monitoring, Green vegetation cover, Leaf area index, PROBA-V, SPOT/VGT}
}

@ARTICLE{usoskin_Icarus_15,
       author = {{Usoskin}, I. and {Kovaltsov}, G.},
        title = "{The carbon-14 spike in the 8th century was not caused by a cometary impact on Earth}",
      journal = {Icarus},
     keywords = {Comets, Cosmic rays, Solar wind, Astrophysics - Earth and Planetary Astrophysics},
         year = 2015,
       volume = {260},
        pages = {475-476},
          %doi = {10.1016/j.icarus.2014.06.009},
archivePrefix = {arXiv},
       eprint = {1401.5945},
 primaryClass = {astro-ph.EP},
       adsurl = {https://ui.adsabs.harvard.edu/abs/2015Icar..260..475U},
      adsnote = {Provided by the SAO/NASA Astrophysics Data System}
}

@article{Koldobskiy2023,
    title = {{Multiproxy Reconstructions of Integral Energy Spectra for Extreme Solar Particle Events of 7176 BCE, 660 BCE, 775 CE, and 994 CE}},
    year = {2023},
    journal = {Journal of Geophysical Research: Space Physics},
    author = {Koldobskiy, Sergey and Mekhaldi, Florian and Kovaltsov, Gennady and Usoskin, Ilya},
    number = {3},
    month = {3},
    pages = {e2022JA031186},
    volume = {128},
    url = {https://onlinelibrary.wiley.com/doi/10.1029/2022JA031186},
    %doi = {10.1029/2022JA031186},
    issn = {2169-9380}
}

@article{OHare2019,
    title = {{Multiradionuclide evidence for an extreme solar proton event around 2,610 B.P. (660 BC)}},
    year = {2019},
    journal = {Proceedings of the National Academy of Sciences of the United States of America},
    author = {O’Hare, Paschal and Mekhaldi, Florian and Adolphi, Florian and Raisbeck, Grant and Aldahan, Ala and Anderberg, Emma and Beer, Jürg and Christl, Marcus and Fahrni, Simon and Synal, Hans Arno and Park, Junghun and Possnert, Göran and Southon, John and Bard, Edouard and Muscheler, Raimund},
    number = {13},
    pages = {5961--5966},
    volume = {116},
    %doi = {10.1073/pnas.1815725116},
    issn = {10916490},
    pmid = {30858311},
    keywords = {Ice cores, Radionuclides, Solar proton events, Solar storms}
}

@article{Mekhaldi2015,
    title = {{Multiradionuclide evidence for the solar origin of the cosmic-ray events of AD 774/5 and 993/4}},
    year = {2015},
    journal = {Nature Communications},
    author = {Mekhaldi, Florian and Muscheler, Raimund and Adolphi, Florian and Aldahan, Ala and Beer, Jürg and McConnell, Joseph R. and Possnert, Göran and Sigl, Michael and Svensson, Anders and Synal, Hans-Arno and Welten, Kees C. and Woodruff, Thomas E.},
    number = {1},
    month = {12},
    pages = {8611},
    volume = {6},
    url = {http://www.nature.com/articles/ncomms9611},
    %doi = {10.1038/ncomms9611},
    issn = {2041-1723}
}

@article{Koldobskiy2021,
    title = {{New reconstruction of event-integrated spectra (spectral fluences) for major solar energetic particle events}},
    year = {2021},
    journal = {Astronomy {\&} Astrophysics},
    author = {Koldobskiy, S. and Raukunen, O. and Vainio, R. and Kovaltsov, G. A. and Usoskin, I.},
    month = {3},
    pages = {A132},
    volume = {647},
    url = {https://www.aanda.org/10.1051/0004-6361/202040058},
    %doi = {10.1051/0004-6361/202040058},
    issn = {0004-6361},
    arxivId = {2101.10234},
    keywords = {Sun: activity, Sun: flares, Sun: particle emission, activity, flares, particle emission, solar-terrestrial relations, sun}
}

@article{Golubenko2025,
    title = {{New SOCOL:14C-Ex model reveals that the Late-Glacial radiocarbon spike in 12350 BC was caused by the record-strong extreme solar storm}},
    year = {2025},
    journal = {Earth and Planetary Science Letters},
    author = {Golubenko, Kseniia and Usoskin, Ilya and Rozanov, Eugene and Bard, Edouard},
    pages = {119383},
    volume = {661},
    publisher = {Elsevier B.V.},
    url = {https://doi.org/10.1016/j.epsl.2025.119383 https://linkinghub.elsevier.com/retrieve/pii/S0012821X25001827},
    %doi = {10.1016/j.epsl.2025.119383},
    issn = {0012821X},
    keywords = {CCM SOCOL:14C-Ex, Extreme solar particle events, Glacial epoch, Radiocarbon, extreme solar particle events}
}

@article{Poluianov2016,
    title = {{Production of cosmogenic isotopes 7 Be, 10 Be, 14 C, 22 Na, and 36 Cl in the atmosphere: Altitudinal profiles of yield functions}},
    year = {2016},
    journal = {Journal of Geophysical Research: Atmospheres},
    author = {Poluianov, S V and Kovaltsov, G A and Mishev, A L and Usoskin, I G},
    number = {13},
    month = {7},
    pages = {8125--8136},
    volume = {121},
    url = {http://doi.wiley.com/10.1002/2016JD025034},
    %doi = {10.1002/2016JD025034},
    issn = {2169897X},
    keywords = {10.1002/2016JD025034 and cosmogenic radionuclides, atmospheric cascade, cosmic ray}
}

@article{Sakurai2020,
    title = {{Prolonged production of 14C during the 660 BCE solar proton event from Japanese tree rings}},
    year = {2020},
    journal = {Scientific Reports},
    author = {Sakurai, Hirohisa and Tokanai, Fuyuki and Miyake, Fusa and Horiuchi, Kazuho and Masuda, Kimiaki and Miyahara, Hiroko and Ohyama, Motonari and Sakamoto, Minoru and Mitsutani, Takumi and Moriya, Toru},
    number = {1},
    month = {1},
    pages = {660},
    volume = {10},
    url = {https://www.nature.com/articles/s41598-019-57273-2},
    %doi = {10.1038/s41598-019-57273-2},
    issn = {2045-2322},
    pmid = {31959822}
}

@article{Guttler2015,
    title = {{Rapid increase in cosmogenic 14C in AD 775 measured in New Zealand kauri trees indicates short-lived increase in 14C production spanning both hemispheres}},
    year = {2015},
    journal = {Earth and Planetary Science Letters},
    author = {G{\"{u}}ttler, D and Adolphi, F and Beer, J and Bleicher, N and Boswijk, G and Christl, M and Hogg, A and Palmer, J and Vockenhuber, C and Wacker, L and Wunder, J},
    pages = {290--297},
    volume = {411},
    url = {https://www.sciencedirect.com/science/article/pii/S0012821X14007481},
    %doi = {https://doi.org/10.1016/j.epsl.2014.11.048},
    issn = {0012-821X},
    keywords = {AD 775, AMS, C, carbon cycle, modeling}
}

@article{Park2017,
    title = {{Relationship between solar activity and {$\Delta$} 14 C peaks in AD 775, AD 994, and 660 BC}},
    year = {2017},
    journal = {Radiocarbon},
    author = {Park, Junghun and Southon, John and Fahrni, Simon and Creasman, Pearce Paul and Mewaldt, Richard},
    number = {4},
    month = {8},
    pages = {1147--1156},
    volume = {59},
    url = {https://www.cambridge.org/core/product/identifier/S0033822217000595/type/journal_article},
    %doi = {10.1017/RDC.2017.59},
    issn = {0033-8222},
    keywords = {660 BC, AD 775, CME (coronal mass ejection), M12, solar energetic particles (SPE)}
}

@ARTICLE{usoskin_1956_20,
       author = {{Usoskin}, I.G. and {Koldobskiy}, S.A. and {Kovaltsov}, G.A. and {Rozanov}, E.V. and
         {Sukhodolov}, T.V. and {Mishev}, A.L. and {Mironova}, I.A.},
        title = "{Revisited Reference Solar Proton Event of 23 February 1956: Assessment of the Cosmogenic-Isotope Method Sensitivity to Extreme Solar Events}",
      journal = {J. Geophys. Res. (Space Phys.)},
     keywords = {solar energetic particles, extreme events, cosmogenic isotopes, Astrophysics - Solar and Stellar Astrophysics, Astrophysics - Earth and Planetary Astrophysics, Physics - Space Physics},
         year = 2020,
       volume = {125},
       number = {6},
          eid = {e27921},
        pages = "{e27921}",
          %doi = {10.1029/2020JA027921},
archivePrefix = {arXiv},
       eprint = {2005.10597},
 primaryClass = {astro-ph.SR},
       adsurl = {https://ui.adsabs.harvard.edu/abs/2020JGRA..12527921U},
      adsnote = {Provided by the SAO/NASA Astrophysics Data System}
}

@article{Uusitalo2018,
    title = {{Solar superstorm of AD 774 recorded subannually by Arctic tree rings}},
    year = {2018},
    journal = {Nature Communications},
    author = {Uusitalo, J and Arppe, L and Hackman, T and Helama, S and Kovaltsov, G and Mielik{\"{a}}inen, K and M{\"{a}}kinen, H and N{\"{o}}jd, P and Palonen, V and Usoskin, I and Oinonen, M},
    number = {1},
    month = {8},
    pages = {3495},
    volume = {9},
    url = {https://doi.org/10.1038/s41467-018-05883-1 https://www.nature.com/articles/s41467-018-05883-1},
    %doi = {10.1038/s41467-018-05883-1},
    issn = {2041-1723}
}

@article{Usoskin2013,
    title = {{The AD775 cosmic event revisited: the Sun is to blame}},
    year = {2013},
    journal = {Astronomy {\&} Astrophysics},
    author = {Usoskin, I. G. and Kromer, B. and Ludlow, F. and Beer, J. and Friedrich, M. and Kovaltsov, G. A. and Solanki, S. K. and Wacker, L.},
    month = {4},
    pages = {L3},
    volume = {552},
    url = {http://www.aanda.org/10.1051/0004-6361/201321080},
    %doi = {10.1051/0004-6361/201321080},
    issn = {0004-6361},
    keywords = {Sun: activity, Sun: flares}
}

@article{Mekhaldi2021,
    title = {{The Signal of Solar Storms Embedded in Cosmogenic Radionuclides: Detectability and Uncertainties}},
    year = {2021},
    journal = {Journal of Geophysical Research: Space Physics},
    author = {Mekhaldi, F. and Adolphi, F. and Herbst, K. and Muscheler, R.},
    number = {8},
    month = {8},
    pages = {e2021JA029351},
    volume = {126},
    url = {https://onlinelibrary.wiley.com/doi/10.1029/2021JA029351},
    %doi = {10.1029/2021JA029351},
    issn = {2169-9380},
    keywords = {cosmogenic radionuclides, ice cores, solar particle events, solar storms}
}

@article{Stenke2013,
    title = {{The SOCOL version 3.0 chemistry–climate model: description, evaluation, and implications from an advanced transport algorithm}},
    year = {2013},
    journal = {Geoscientific Model Development},
    author = {Stenke, A and Schraner, M and Rozanov, E and Egorova, T and Luo, B and Peter, T},
    number = {5},
    pages = {1407--1427},
    volume = {6},
    url = {https://gmd.copernicus.org/articles/6/1407/2013/},
    %doi = {10.5194/gmd-6-1407-2013}
}

@article{Brehm2025,
    title = {{Tracing ancient solar cycles with tree rings and radiocarbon in the first millennium BCE}},
    year = {2025},
    journal = {Nature Communications },
    author = {Brehm, Nicolas and Pearson, Charlotte L. and Christl, Marcus and Bayliss, Alex and Nicolussi, Kurt and Pichler, Thomas and Brown, David and Wacker, Lukas},
    number = {1},
    pages = {1--10},
    volume = {16},
    publisher = {Springer US},
    url = {http://dx.doi.org/10.1038/s41467-024-55757-y},
    %doi = {10.1038/s41467-024-55757-y},
    issn = {20411723}
}

@article{Uusitalo2024,
    title = {{Transient Offset in 14 C After the Carrington Event Recorded by Polar Tree Rings}},
    year = {2024},
    journal = {Geophysical Research Letters},
    author = {Uusitalo, Joonas and Golubenko, Kseniia and Arppe, Laura and Brehm, Nicolas and Hackman, Thomas and Hayakawa, Hisashi and Helama, Samuli and Mizohata, Kenichiro and Miyake, Fusa and M{\"{a}}kinen, Harri and N{\"{o}}jd, Pekka and Tanskanen, Eija and Tokanai, Fuyuki and Rozanov, Eugene and Wacker, Lukas and Usoskin, Ilya and Oinonen, Markku},
    number = {5},
    month = {3},
    pages = {e2023GL106632},
    volume = {51},
    url = {https://agupubs.onlinelibrary.wiley.com/doi/10.1029/2023GL106632},
    %doi = {10.1029/2023GL106632},
    issn = {0094-8276}
}

@article{Buntgen2018,
    title = {{Tree rings reveal globally coherent signature of cosmogenic radiocarbon events in 774 and 993 CE}},
    year = {2018},
    journal = {Nature Communications},
    author = {B{\"{u}}ntgen, Ulf and Wacker, Lukas and Galv{\'{a}}n, J. Diego and Arnold, Stephanie and Arseneault, Dominique and Baillie, Michael and Beer, Jürg and Bernabei, Mauro and Bleicher, Niels and Boswijk, Gretel and Br{\"{a}}uning, Achim and Carrer, Marco and Ljungqvist, Fredrik Charpentier and Cherubini, Paolo and Christl, Marcus and Christie, Duncan A. and Clark, Peter W. and Cook, Edward R. and D’Arrigo, Rosanne and Davi, Nicole and Eggertsson, Ólafur and Esper, Jan and Fowler, Anthony M. and Gedalof, Ze’ev and Gennaretti, Fabio and Grie{\ss}inger, Jussi and Grissino-Mayer, Henri and Grudd, Håkan and Gunnarson, Björn E. and Hantemirov, Rashit and Herzig, Franz and Hessl, Amy and Heussner, Karl-Uwe and Jull, A. J. Timothy and Kukarskih, Vladimir and Kirdyanov, Alexander and Kol{\'{a}}{\v{r}}, Tomáš and Krusic, Paul J. and Kyncl, Tomáš and Lara, Antonio and LeQuesne, Carlos and Linderholm, Hans W. and Loader, Neil J. and Luckman, Brian and Miyake, Fusa and Myglan, Vladimir S. and Nicolussi, Kurt and Oppenheimer, Clive and Palmer, Jonathan and Panyushkina, Irina and Pederson, Neil and Rybn{\'{i}}{\v{c}}ek, Michal and Schweingruber, Fritz H. and Seim, Andrea and Sigl, Michael and Churakova, Olga and Speer, James H. and Synal, Hans-Arno and Tegel, Willy and Treydte, Kerstin and Villalba, Ricardo and Wiles, Greg and Wilson, Rob and Winship, Lawrence J. and Wunder, Jan and Yang, Bao and Young, Giles H. F.},
    number = {1},
    month = {12},
    pages = {3605},
    volume = {9},
    url = {http://www.nature.com/articles/s41467-018-06036-0},
    %doi = {10.1038/s41467-018-06036-0},
    issn = {2041-1723},
    pmid = {30190505}
}

@article{Brehm2022,
    title = {{Tree-rings reveal two strong solar proton events in 7176 and 5259 BCE}},
    year = {2022},
    journal = {Nature Communications},
    author = {Brehm, Nicolas and Christl, Marcus and Knowles, Timothy D J and Casanova, Emmanuelle and Evershed, Richard P and Adolphi, Florian and Muscheler, Raimund and Synal, Hans-Arno and Mekhaldi, Florian and Paleari, Chiara I and Leuschner, Hanns-Hubert and Bayliss, Alex and Nicolussi, Kurt and Pichler, Thomas and Schl{\"{u}}chter, Christian and Pearson, Charlotte L and Salzer, Matthew W and Fonti, Patrick and Nievergelt, Daniel and Hantemirov, Rashit and Brown, David M and Usoskin, Ilya and Wacker, Lukas},
    number = {1},
    month = {3},
    pages = {1196},
    volume = {13},
    url = {https://doi.org/10.1038/s41467-022-28804-9 https://www.nature.com/articles/s41467-022-28804-9},
    %doi = {10.1038/s41467-022-28804-9},
    issn = {2041-1723}
}

@article{Golubenko2022,
    title = {{Zonal Mean Distribution of Cosmogenic Isotope ( 7 Be, 10 Be, 14 C, and 36 Cl) Production in Stratosphere and Troposphere}},
    year = {2022},
    journal = {Journal of Geophysical Research: Atmospheres},
    author = {Golubenko, K and Rozanov, E and Kovaltsov, G and Usoskin, I},
    number = {16},
    month = {8},
    pages = {e2022JD036726},
    volume = {127},
    url = {https://onlinelibrary.wiley.com/doi/10.1029/2022JD036726},
    %doi = {10.1029/2022JD036726},
    issn = {2169-897X}
}

@article{Panyushkina2024,
  author    = {Panyushkina, Irina P. and Jull, A. J. Timothy and Molnár, Mihály and others},
  title     = {The timing of the ca-660 BCE Miyake solar-proton event constrained to between 664 and 663 BCE},
  journal   = {Communications Earth \& Environment},
  year      = {2024},
  volume    = {5},
  pages     = {454},
  %doi       = {10.1038/s43247-024-01618-x},
  url       = {https://doi.org/10.1038/s43247-024-01618-x}
}

\end{document}